\documentclass[a4paper,fleqn,usenatbib]{mnras}

\usepackage{txfonts}

\usepackage[T1]{fontenc}
\usepackage{ae,aecompl}

\usepackage{graphicx}	
\usepackage{amstext}	
\usepackage{amssymb}	
 
\title[Probing IGM parameters using QSOs]{Probing the parameters of the intergalactic medium using quasars}

\author[T. Dalton et al]{
Tony Dalton,$^{1}$\thanks{E-mail:tonydalton@live.ie}
Simon L. Morris,$^{1}$
Michele Fumagalli,$^{2,3}$
and Efrain Gatuzz$^{4}$\\
$^{1}$Centre for Extragalactic Astronomy, Durham University, South Road, Durham DH1 3LE, UK\\
$^{2}$Dipartimento di Fisica `G. Occhialini', Universit\`a degli Studi di Milano-Bicocca, Piazza della Scienza 3, 20126 Milano, Italy\\
$^{3}$INAF - Osservatorio Astronomico di Trieste, via G. B. Tiepolo 11, 34143 Trieste, Italy \\
$^{4}$Max-Planck-Institut f\"ur extraterrestrische Physik, Gie{\ss}enbachstra{\ss}e 1, 85748 Garching, Germany
\\
}

\date{Accepted XXX. Received YYY; in original form ZZZ}

\pubyear{2015}

\begin{document}
\label{firstpage}
\pagerange{\pageref{firstpage}--\pageref{lastpage}}
\maketitle

\begin{abstract}
We continue our series of papers on intergalactic medium (IGM) tracers using quasi-stellar objects (QSOs), having examined gamma-ray bursts (GRBs) and blazars in earlier studies. We have estimated  the IGM properties of hydrogen column density ($\mathit{N}\textsc{hxigm}$), temperature and metallicity using \textit{XMM-Newton} QSO spectra over a large redshift range,  with a collisional ionisation equilibrium (CIE)  model for the ionised plasma.  The $\mathit{N}\textsc{hxigm}$ parameter results were robust with respect to intrinsic power laws, spectral counts, reflection hump and soft excess features. There is scope for a luminosity bias given both luminosity and $\mathit{N}\textsc{hxigm}$ scale with redshift, but we find this unlikely given the consistent IGM parameter results across the other tracer types reviewed. The impact of intervening high column density absorbers was found to be minimal. The $\mathit{N}\textsc{hxigm}$ from the QSO sample scales as 
$(1 + z)^{1.5\pm0.2}$.  
The mean hydrogen density at $z = 0$  is $n_0 = (2.8\pm{0.3}) \times 10^{-7}$ cm$^{-3}$, the mean IGM temperature over the full redshift range is log($T$/K) $= 6.5\pm{0.1}$, and the mean metallicity is $[X/$H$] = -1.3\pm0.1 (Z \sim 0.05)$. Aggregating with our previous GRB and blazar tracers, we conclude that we have provided evidence of the IGM contributing substantially and consistently to the total X-ray absorption seen in the spectra. These results are based on the necessarily simplistic slab model used for the IGM, due to the inability of current X-ray data to constrain the IGM redshift distribution.
\end{abstract}




\section{Introduction}\label{introduction}

Most baryonic matter in the IGM is not in the form of luminous virialized matter \citep{Shull2012}, and the majority of hydrogen and helium is ionized. In order to measure the IGM density, metallicity and temperature, the observation of metals is essential.  Powerful cosmological sources such as GRBs, blazars and QSOs are currently the most effective targets to study the IGM out to high redshift as their X-ray absorption provides information on the total absorbing column density of the matter, subject to the IGM model chosen and assumptions.  

QSOs are an extremely luminous form of Active Galactic Nuclei (AGN) observed over a huge cosmological range with the current most distant being J0313-1806 at z = 7.642 \citep{Wang2020}. Under the generally accepted scenario, UV emission in QSOs is produced by viscous dissipation in the accretion disk where the gravitational energy of the infalling material is partially transformed into radiation \citep{shakura1973}. The UV photons are Comptonised to X-rays by a corona of hot relativistic electrons around the accretion disk \citep{Haardt1993}. These X-rays can illuminate the accretion disc, being reflected back towards the observer. The observational signs of such reflection features are iron emission lines, Fe K absorption edge and Compton scattering hump. However, these are not always apparent or observed. While features such as the Compton hump, soft excess and iron emission lines are frequently observed in lower luminosity AGN, particularly at lower redshift, they are not often observed in QSOs where the very powerful emission continuum dominates \citep[and references therein]{Scott2011a}. 
QSOs have been extensively studied for many decades across a very wide band of frequencies from radio to X-ray. The availability of ultra-violet (UV) databases and catalogues enables broadband comparison with X-rays for our purposes. The clear non-linear relation between the UV and X-ray components has been measured in detail, and noted to be reasonably constant over redshift and luminosity ranges \citep[e.g.][and references therein]{Risaliti2019,Salvestrini2019,Lusso2020}. The very consistent spectra of QSOs observed over an extensive redshift range make them attractive as IGM tracers, as it can then be hypothesised that deficits or hardening in continuum curvature that are related to redshift could be interpreted as signatures of IGM absorption.  

QSOs as X-ray tracers of the IGM have been well studied in the past \citep[e.g.][]{Wilkes1987,Elvis1994a,Page2005,Behar2011,Starling2013}. X-ray absorption is typically dominated by metal ions and reported as an equivalent hydrogen column density ($\mathit{N}\textsc{hx}$). The early observations of excess absorption in QSOs at high redshift in X-ray over the known Galactic absorption  ($\mathit{N}\textsc{hxGal}$) were unexpected, as in X-ray, the absorbing cross-section decreases as the observed spectral energy increases with redshift \citep[e.g.][]{Elvis1994a,Cappi1997, Fiore1998,Elvis1998}. This excess absorption was initially assumed to be located in the QSO host. \citet{Reeves2000} were among the first to strongly advocate a relation between excess absorption and redshift but noted that the assumption of all such excess being at the QSO rest-frame could lead to overestimation of column densities as the absorbing material could lie anywhere on the LOS. Later studies explored the possibility of the IGM contributing to the excess absorption and found it to be related to redshift \citep[e.g.][]{Eitan2013,Starling2013,Arcodia2018}. However, all such studies assumed by convention that the absorbers were neutral and at solar metallicity. As typical QSO hosts, and IGM absorbers are partially ionised and have low metallicity, the resulting reported column densities are, therefore, lower limits. In our previous studies on GRBs \citep[hereafter D20 and D21a]{Dalton2020,Dalton2021a} and blazars \citep[hereafter D21b]{Dalton2021b}, we used realistic parameter ranges for metallicity and temperature in collisional ionisation absorption models for the IGM. We found strong evidence for IGM X-ray mean column density rising with redshift in the spectra of both GRBs and blazars. We now continue the series using similar IGM and continuum models to study QSO spectra. In this paper, all data are taken from the European Space Agency’s $\mathit{XMM-Newton}$ Photon Imaging camera (EPIC) \citep{Struder2001} which has reasonable response down to 0.15 keV, high sensitivity to extended emission, and large effective area enabling detailed analysis of soft X-ray properties. $\mathit{XMM-Newton}$ has three cameras, PN, MOS-1 and MOS-2. Our data are taken from PN except for our highest redshift QSOs where we included the MOS-1 and MOS-2 data to increase spectral counts.

In our previous papers in this series, we studied GRBs and blazars as tracers of IGM properties and possible variation with redshift (D20, D21a and D21b). We continue the series in this paper with the study of QSOs. Our main objective is to estimate the IGM column density, temperature and metallicity, using an ionised absorption model, on the line of sight (LOS) to QSOs. Our continuing hypothesis is that the integrated IGM column density from IGM absorption increases with redshift.  We analyse this highly ionised IGM absorption in addition to examining appropriate host environment and continuum intrinsic models. We test the robustness of our results and aggregate our QSO sample with our GRB and blazar samples for cross-tracer comparison.

The sections that follow are: Section \ref{sec:data} describes the data selection and methodology; Section \ref{sec:models} covers the models for the IGM LOS including assumptions and parameters, and QSO continuum models; Section \ref{sec:QSO results} gives the results of QSO spectra fits using collisional IGM models with free IGM key parameters; in Section \ref{sec:Robust} we test the robustness of the IGM model fits including a review of the QSO UV spectra for any high density absorbers; in Section \ref{sec:combinedQSOGRBblazar} we aggregate GRB and blazar samples with our QSO sample for cross-tracer analysis. In Section \ref{sec:Discuss}, we discuss and compare results with other studies and Section \ref{sec:conclusion} gives conclusions. We suggest readers interested in the IGM property results see Sections \ref{sec:QSO results}, \ref{sec:combinedQSOGRBblazar} and \ref{sec:conclusion}. For spectra fitting methodology and model assumptions readers should also go to Sections \ref{sec:data} and \ref{sec:models}. Finally, for more detailed examination of robustness of the QSO spectra fitting and discussion on other studies, read Sections \ref{sec:Robust} and \ref{sec:Discuss}. In this paper where relevant, we adopt the cosmological parameters $\Omega_M = 0.3$, $\Omega_\Lambda = 0.7$, and $H_0 = 70$ km s$^{-1}$ Mpc$^{-1}$.

\section{Data selection and methodology}\label{sec:data}

\begin{table*}
    \renewcommand{\arraystretch}{1.2}
	\centering
	\caption{SDSS-DR14 and $\textit{4XMM-Newton-DR9}$ cross-correlation QSO sample. For each QSO, the columns give the name, radio type (radio loud - RLQ, radio quiet - RQQ, or unknown), redshift, number of counts in 0.3-10 keV range, count rate ($s^{-1}$), Galactic column density (log($\mathit{N}\textsc{hxGal}/$cm$^{-2}$)) and unabsorbed luminosity (2-10keV)(log(L/erg $s^{-1}$)). Co-added spectra for a number of QSOs are used, often observed over a period of time, so we do not provide individual observation information.}
	\label{tab:Table_fullsample}
	\begin{tabular}{cc@{\hspace*{0.5cm}} c@{\hspace*{0.8
	cm}}c@{\hspace*{0.8
	cm}}c@{\hspace*{0.4cm}}c@{\hspace*{0.4cm}}c@{\hspace*{0.4
	cm}}}
		\hline
		
	    QSO & Radio type & $z$ & Total counts & Mean count  & log & log \\
		
		& & & & rate($s^{-1}$& ($\mathit{N}\textsc{hxGal}/$cm$^{-2}$) & (L/erg $s^{-1}$)\\
		\hline

        J142952+544717* & RLQ& 6.18 & 725& 0.046  & 20.18 & 46.36\\
		022112.62-034252.2* & unknown & 5.01 & 339 &0.034 &20.30  & 45.14  \\
		001115.23+144601.8 & RLQ& 4.96 & 2258& 0.096  &20.31 & 46.47\\
		143023.73+420436.5 & RLQ & 4.71 & 13162 &0.157 & 20.29&47.11  \\
		223953.6-055220.0 & RQQ & 4.56 & 450 &0.015 &20.58 &45.63 \\
        151002.93+570243.3 & RLQ & 4.31 & 1395 &0.15 &20.17 &45.88  \\
        
        133529.45+410125.9& RQQ & 4.26 & 626 &0.055 &19.98 & 46.09  \\
		132611.84+074358.3 & RQQ & 4.12 & 947 &0.025 & 20.48&46.09 \\
		163950.52+434003.7& RLQ & 3.99 & 1158 &0.029 & 20.30 & 45.76 \\
		021429.29-051744.8& RLQ & 3.98 & 1126 &0.018 &20.30 & 45.57\\
        133223.26+503431.3 & RQQ & 3.81 & 404 &0.022 & 20.03 & 45.60  \\
        200324.1-325144.0 &RLQ & 3.78 & 3484 &0.23 & 20.86 & 46.56 \\
        200324.1-135245.1 & RLQ  & 3.77 & 2963 &0.21 & 20.90 & 45.86 \\
		
        122135.6+280614.0 & RLQ & 3.31 & 2994 &0.093 & 20.30 & 45.35 \\
        042214.8-384453.0 & RLQ & 3.11 & 1840 &0.22 &20.31 & 45.48 \\
        083910.89+200207.3& RLQ & 3.03 & 4251 &0.103 & 20.3 & 45.86 \\
		
        111038.64+483115.6&RQQ & 2.96 & 741 &0.022 & 20.10 & 45.35
        \\
        
        122307.52+103448.2 & RQQ& 2.75 & 535 &0.029 & 20.35 & 45.48 \\
		115005.36+013850.7& unknown& 2.33 & 954 &0.014 & 20.36 & 45.19 \\
		121423.02+024252.8 &RLQ & 2.22 & 5394 &0.077 & 20.25 & 45.62 \\
        112338.14+052038.5 &RLQ & 2.18 & 826 &0.031 & 20.64 & 45.62 \\
        123527.36+392824.0 & RQQ& 2.16 & 553 &0.017 & 20.17 & 45.07\\
        134740.99+581242.2&RLQ & 2.05 & 2978&0.112 & 20.11 & 45.66\\
		095834.04+024427.1& RQQ& 1.89 & 1444 &0.023 & 20.44 & 44.87 \\
        093359.34+551550.7 & RQQ  & 1.86 &2309 &0.09 & 20.26 & 45.64 \\
        133526.73+405957.5 &RQQ & 1.77 & 634 &0.062 &19.97 & 45.39 \\
        100434.91+411242.8&RQQ & 1.74 & 9558 &0.27 & 20.05 & 45.90\\
		104039.54+061521.5 &RLQ & 1.58 & 946
		&0.019 & 20.45 & 44.94 \\
        083205.95+524359.3 &RQQ & 1.57 & 1303 &0.016  & 20.58 & 44.61  \\
        112320.73+013747.4 &RQQ & 1.47 & 1801 &0.078 & 20.62 & 45.37 \\
        091301.03+525928.9&RQQ & 1.38 & 1221 &0.44 & 20.20 & 45.88\\
		121426.52+140258.9 &RLQ & 1.28 & 946&0.019 & 20.44 & 45.19 \\
        105316.75+573550.8 &RQQ & 1.21 & 2059 &0.066 & 19.75 & 45.12\\
        085808.91+274522.7 &RQQ & 1.09 & 3158 &0.043  & 20.49 & 44.71\\
        095857.34+021314.5&RQQ & 1.02 & 1904 &0.77 & 20.43 & 45.10 \\
		125849.83-014303.3 &RQQ & 0.97& 7032&0.20  & 20.20 & 45.09\\
        082257.55+404149.7&RLQ & 0.86 & 815 &0.158 & 20.65 & 44.96\\
        150431.30+474151.2 &RQQ & 0.82 & 1499 &0.106 & 20.34 & 45.02\\
        111606.97+423645.4&RQQ & 0.67& 2409 &0.081 & 20.25 & 44.57\\
		130028.53+283010.1 &RLQ & 0.65& 6859&0.314 & 19.97 & 45.03 \\
        111135.76+482945.3&RQQ & 0.56 & 4081 &0.150 & 20.10 & 44.67\\
        091029.03+542719.0&RQQ & 0.53 & 2073 &0.058 & 29.32 & 44.16\\
		105224.94+441505.2 &RQQ & 0.44& 1237&0.156 & 20.05 & 44.19 \\
        223607.68+134355.3&RQQ & 0.33& 3106 &0.058 & 20.68 & 44.30\\
        144645.93+403505.7 &RQQ & 0.27 & 15843 &0.959 & 20.10 & 44.14\\
        123054.11+110011.2&RQQ & 0.24& 6368 &1.158 & 20.33 & 44.33\\
		103059.09+310255.8 &RLQ & 0.18& 37274&1.79 & 20.29 & 44.36\\
        141700.81+445606.3&RQQ & 0.11 & 29070 &1.386 & 20.09 & 43.56\\
    \hline
    \end{tabular}
    $^*\textrm{These QSOs had poor high energy spectra above 2keV so the range taken was from 0.2 - 2.0 keV}$	

\end{table*}

Our sample of QSOs is taken from the catalogue created by \citet{Lusso2020} based on the 14th Data Release of the \textit{Sloan Digital Sky Survey} (SDSS-DR14) \citep{York2000}  which they cross-matched with 4\textit{XMM-Newton} Data Release-9 data giving an initial sample of 24,947 QSOs. We applied an initial minimum threshold of X-ray counts $> 500$ for the PN camera to ensure high signal-to-noise spectra. As the number of QSOs with $z > 4$ decreases dramatically, we drew from samples in \citet{Page2005,Grupe2006,Eitan2013,Nanni2017,Vito2019,Medvedev2021}. For $z < 4$ QSOs, we selected those with highest counts, maintaining a redshift spread. We relaxed our minimum count cutoff requirement slightly above $z \sim3.8$, with 3 QSOs have counts between $400 - 500$. The highest two redshift QSOs have data from all three EPIC cameras to increase the spectral counts. Our final sample of 48 QSOs has a redshift range of $0.114 \leq z \leq 6.18$ 

Radio loudness (R) is typically defined as the ratio of the flux densities at rest-frame 5GHz and 4400 \r{A}, with R $\geq 10$ and R  < 10 for radio-loud (RLQ) and radio-quiet (RQQ) respectively \citep{Kellermann1989}. We include both RLQ and RQQ in our sample but exclude broad absorption line QSOs as these are known to be highly absorbed in X-ray and could dominate any possible IGM absorption. In general, for a given optical luminosity, the X-ray emission from RLQs is about three times greater than that from RQQs  which allows them to be studied out to higher redshifts \citep[and references therein]{Scott2011a}. As a result, 19 out of our 48 QSOs are RLQ which may be a source of bias given on average, approximately $10\%$ of QSOs are RLQ \citep[e.g][]{Grupe2006}. We explore this in Section \ref{sec:Robust}.

The $\textit{XMM-Newton}$ EPIC spectra were obtained in timing mode and reduced with the Science Analysis System (SAS2, version 19.1.0). First, we processed each observation with the \textsc{epchain} SAS tool. We used only single-pixel events (PATTERN==0) while bad time intervals were filtered out for large flares applying a 1.0 cts s$^{-1}$ threshold. In order to avoid bad pixels and regions close to CCD edges, we filtered the data using FLAG==0. We manually inspected the source and background subtraction region for each observation. 

For our fitting, we use \textsc{xspec} version 12.11.1 \citep{Arnaud1996}. We use the C-statistic \citep{Cash1979} (Cstat) which is based on the Poisson likelihood and gives more reliable results for small number counts per bin.   As we are using total X-ray spectral absorption for the IGM, we can expect some degeneracy between the parameters. We, therefore, follow the same method as in our other papers in this series (D21a and D21b) using both \textsc{steppar} function and Markov chain Monte Carlo to overcome the problem of local probability maxima, and to give confidence intervals on our IGM property results. We adopt the approach that a reduction of Cstat$> 2.71$, $>4.6$, and $>6.25$ for one, two and three additional interesting parameters corresponds to $90\%$  significance \citep{Reeves2000,Ricci2017}. To  avoid empty channels, we binned spectra to have a minimum count of one count per bin so the Cstat value is independent of the count numbers \citep{Nanni2017}.
We assume a homogeneous isotropic IGM as all our QSO sample have LOS much greater than the large scale structure, while acknowledging that large individual absorbers can still impact the LOS (tested in Section \ref{sec:Robust}).

\section{Models for the QSO continuum and LOS features}\label{sec:models}

In this section, we describe our IGM models and parameter ranges, the models used for fitting the intrinsic spectra, absorption of the QSOs, and our Galaxy. We emphasise that we are not attempting to find a model fully consistent with the QSO spectrum, so long as our intrinsic model sufficiently represents spectral curvature and shape, with the remaining spectral features being attributable to the IGM. We, therefore, do not necessarily expect our modelling to yield any physical insight into the nature of the QSO engine itself. Given the moderate resolution of \textit{XMM-Newton}, our spectral modelling and analysis pertains to the overall continuum absorption and not individual lines, edges or features (D21a and b). 

\subsection{Galactic absorption}
We use \textsc{tbabs} \citep[hereafter W00]{Wilms2000} with $\textit{N}\textsc{hxGal}$ fixed to the values based on  \citet{Willingale2013} and \citet{Kalberla2005}. We use W00 solar abundances which factor in H$_2$ and dust in the galaxy interstellar medium.  

\subsection{Continuum models}
In the energy range 0.3-10 keV, QSO spectra are typically modelled with a simple power law. Some studies add a high energy cut-off at $\sim100$keV or higher \citep[e.g.][]{Ricci2017}, but such cut-off values are well outside our X-ray energy range. Many QSOs show curvature, particularly in soft X-ray and a log-parabolic power law can be more appropriate. A Compton reflection hump is a common feature in QSOs, mainly RQQ. However, the visibility of this component in the observed spectra of QSOs is low, as their emission is mainly dominated by the luminous continuum  \citep[e.g.][]{Reeves2000,Scott2011a}. In high luminosity QSOs, the reflection component may be intrinsically weaker due to possible ionisation of the inner accretion disc, reducing the neutral matter available to generate a reflection feature \citep{Mushotzky1993}. There is little observational evidence, particularly for higher redshift QSOs ($z >2$) of the iron emission line, probably due to the dominant emission continuum \citep{Page2005}. QSOs sometimes show a soft excess, particularly at lower redshifts. This was initially postulated to be the hard tail of the UV 'big blue bump'. While there is no consensus on the origin of the soft excess, there are now several prominent theories e.g. an artifact of ionised absorption \citep[e.g][]{Gierlinski2004}, Comptonisation of UV photons \citep[e.g.][]{Done2012}, and relativistically blurred disc reflection \citep[e.g.][]{Crummy2006}. As the soft excess rarely shows above redshift $z > 0.3$, and the reflection hump is also rarely seen in QSOs, we omit adding specific components for these features in our initial fitting. 

Accordingly, we model the QSO continuum with a simple and a log-parabolic power law. In Section \ref{sec:Robust}, we robustly explore whether the inclusion of model components for reflection and/or soft excess improves the fit and/or impacts any IGM absorption.

\subsection{QSO host absorption}
As noted in Section \ref{introduction}, by convention many X-ray QSO studies assume any absorption in excess of our Galaxy is due to the host galaxy, with the absorber assumed to be neutral and with solar metallicity. To more accurately isolate any absorption by the QSO host, we base our model on the findings in the Quasar Probing Quasar series \citep[e.g][]{Hennawi2006,Prochaska2009,Hennawi2013,Prochaska2013}. Accordingly, our host model assumes collisionally ionised absorption (CIE) in the circum-galactic medium (CGM) at fixed parameters of log($\textit{N}\textsc{hx}$/cm$^{-2}) = 20$, log(T/K) = 6 and $[X/$H$] = -1$ $(Z/Z\sun = 0.1)$. We use the $\textsc{xspec}$  CIE model $\textsc{hotabs}$ \citep{Kallman2009}. We note that there is evidence of metallicity evolution in QSOs \citep[e.g.][and references therein]{Prochaska2014} but not sufficient to warrant leaving the metallicity parameter variable in the host model. Further, Damped Lyman Alpha Systems (DLAs) have been observed on the LOS to QSOs. However, their very low incidence means they have limited potential impact on most QSO spectra. We examine this further in Section \ref{sec:Robust}. Finally, we note that the incidence of QSOs with significantly reddened optical spectra is rare, indicating that the dust/gas ratio is low \citep{Page2005}. Therefore, we assume there is no dust impact on the assumed host absorption. We note that our choice of QSO host model precludes any significant host X-ray absorption.

\subsection{Ionised IGM component}

We follow the D21a and D21b methodology for the modelling the IGM absorption. We initially fitted a sub-sample of QSOs with both photoionisation and collisional ionisation equilibrium (PIE and CIE respectively) models separately to study these examples. Similar results for $\mathit{N}\textsc{hxigm}$ were obtained for both models, consistent with D21a and D21b. Some combination of CIE and PIE absorption is the most physically plausible scenario for the full LOS. It is not possible to determine which ionisation model is the better single model for the IGM at all redshifts, and we follow  D21b, fitting with the CIE model \textsc{hotabs} only.  As noted in Section \ref{sec:models}, we are modelling and fitting the overall continuum curvature, and not specific absorption features. We note that this gives scope for possible degeneracy to occur. This degeneracy could arise from the relation between column density, temperature and metallicity, but also due to features such as soft excess and reflection humps. We examine the potential impact of such soft excess and reflection components in Section \ref{sec:Robust}.

Our IGM model assumes a plane-parallel uniform slab geometry in ionization and thermal equilibrium to model the IGM LOS \citep[e.g.][]{Savage2014,Khabibullin2019,Lehner2019}. As an approximation of the full LOS IGM absorption, in a homogeneous medium, this slab is located at half the QSO redshift. In Section \ref{subsec:4.2}, we explore the impact of this slab redshift assumption on the resulting $\textit{N}\textsc{hxigm}$. 

We use the same IGM parameter ranges as D21a and D21b for density, temperature and metallicity as summarised in Table \ref{tab:Table_paramaters}. The metallicity range is broad enough to cover the most diffuse low metallicty IGM regions, to the higher metallicity warm-hot IGM (WHIM) based on \citep[e.g][]{Schaye2003,Aguirre2008,Danforth2016,Pratt2018}.

\begin{table}
	\centering
	\caption{Free parameter limits in the IGM model. Continuum parameters are also left free. The fixed parameters are Galactic  log($\textit{N}\textsc{hx}/$cm$^{-2}$), the IGM slab redshift at half the QSO redshift, and the QSO host CGM log($\textit{N}\textsc{hx}/$cm$^{-2}$), temperature and metallicity. }
	\label{tab:Table_paramaters}
	\begin{tabular}{lccr} 
		\hline
		IGM parameter &  range in \textsc{xspec} models\\
		\hline
		column density & $19 \leq$ log($\textit{N}\textsc{hx}/$cm$^{-2}$) $\leq 23$ \\
		temperature\ & $4 \leq$ log($T$/K) $\leq 8$  \\
		metallicity & $-4 \leq [X/$H$] \leq -0.7$ \\
		\hline
	\end{tabular}
\end{table}

\begin{figure}

	\includegraphics[width=\columnwidth]{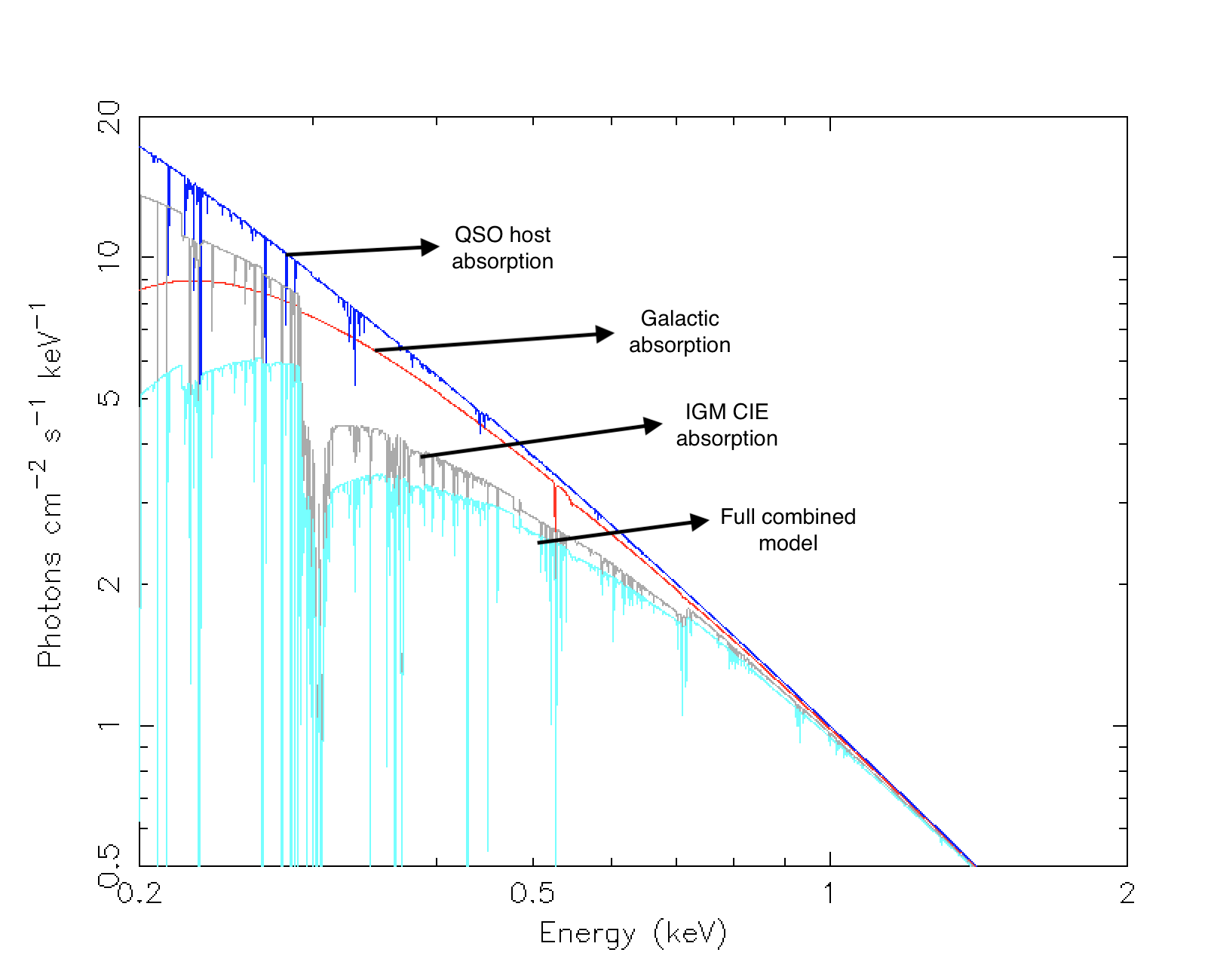}
    \caption{Model components for the LOS absorption to a QSO at $z = 3$, in the energy range 0.2 - 3keV.  Each component is shown separately combined with a log-parabolic power law, as well as the full combined model: IGM CIE absorption (grey) of a slab at $z = 1.5$ log($\textit{N}\textsc{hxigm}/$cm$^{-2}) = 22.00$, $Z = 0.05Z\sun$, and log($T/$K) = 6.00 ; log($\textit{N}\textsc{hx}/$cm$^{-2}) = 20$ for our Galaxy (red); log($\textit{N}\textsc{hx}/$cm$^{-2}) = 20$ with $Z = 0.1Z\sun$ and log($T/$K) = 6.00 for the QSO host CGM at $z = 3$ (blue). The full combined model is the light blue line. }
    \label{fig:Full model}
\end{figure}

Our model components are shown in the example in Fig.\ref{fig:Full model} for the  LOS absorption to a QSO at $z = 3$. We show the model components separately using a log-parabolic power law for each line, as well as the full combined model: CIE IGM absorption (grey) for a slab at $z = 1.5$ log($\textit{N}\textsc{hxigm}/$cm$^{-2}) = 22.00$, $Z = 0.05Z\sun$, and log($T/$K) = 6.00  fo a slab at  $z = 1.5$; log($\textit{N}\textsc{hx}/$cm$^{-2}) = 20$ for our Galaxy (red); log($\textit{N}\textsc{hx}/$cm$^{-2}) = 20$ with $Z = 0.1Z\sun$ and log($T/$K) = 6.00 for the QSO host CGM (blue) at $z = 3$. The full combined model is the light blue line. The absorption lines are clearly visible in the model example, but these features would not be detected in a real spectrum  due to instrument limitations and redshift smearing.

Substantial absorption by intervening neutral absorbers with log($\textit{N}\textsc{h i}/$cm$^{-2}) > 21.00$ is rare in QSO LOS, and insufficient to account for the observed spectral curvature unless there are several intervening DLA or a galaxy  \citep[e.g][]{Elvis1994,Cappi1997,Fabian2001,Page2005}. Accordingly, we omit absorption contribution from any such objects. In Section \ref{sec:Robust}, we will examine all known DLA and intervening galaxies on the QSO LOS to see if they could account for any curvature in the sample spectra.

The full \textsc{xspec} models based on the above components are therefore:\\

\textsc{tbabs}(Galaxy z=0) * \textsc{hotabs}(IGM slab at QSO z/2) * \textsc{hotabs}(host CGM z = zQSO) * \textsc{po}

or

\textsc{tbabs}(Galaxy z=0) * \textsc{hotabs}(IGM slab at QSO z/2) * \textsc{hotabs}(host CGM z = zQSO) * \textsc{logpar}

\section{QSO spectral analysis results}\label{sec:QSO results}

\begin{table*}
    \renewcommand{\arraystretch}{1.3}
	\centering
	\caption{Full IGM model fitting results for the SDSS-DR14 and $\textit{4XMM-Newton-DR9}$ cross-correlation QSO sample. For each QSO, the columns give the name, redshift, IGM paramaters: $\mathit{N}\textsc{hxigm}$, $[X/$H$]$, temperature; Continuum log parabolic power law and curvature parameter $\beta$; Cstat$/$dof}
	\label{tab:Fullresults}
	\begin{tabular}{cc@{\hspace*{0.7cm}} c@{\hspace*{0.5
	cm}}c@{\hspace*{0.5
	cm}}c@{\hspace*{0.5cm}}c@{\hspace*{0.5cm}}c@{\hspace*{0.5cm}}c}
	
	 \hline

		QSO &$z$ & log$(\frac{\mathit{N}\textsc{hxigm}}{\textrm{cm}^{-2}})$ & $[X/\textrm{H}]$ & log$(\frac{\textrm{T}}{ K})$ & PO  & $\beta$ & Cstat$/$dof \\
		\hline

        J142952+544717 & 6.18 & $22.28^{+0.30}_{-0.24}$ &$-1.48^{+0.46}_{-0.44}$  & $5.72^{+0.78}_{-0.39}$ & $2.67^{+0.25}_{-0.26}$& $-0.96^{+1.42}_{-0.02}$  & 483.20/1940 \\
        022112.62-034252.2&5.01 & $22.60^{+0.04}_{-1.03}$ &$-1.14^{+0.11}_{-0.75}$  & $7.91^{+0.04}_{-2.67}$ & $2.13^{+0.37}_{-0.42}$& $0.94^{+0.00}_{-1.65}$  & 110.04/1940
		  \\
		001115.23+144601.8 & 4.96 & $22.47^{+0.12}_{-1.01}$ &$-1.06^{+0.17}_{-1.34}$  & $7.98^{+0.00}_{-2.78}$ & $1.51^{+0.88}_{-0.01}$& $0.64^{+0.22}_{-0.29}$  & 639.73/1940 \\
        143023.73+420436.5 &4.71 & $22.20^{+0.40}_{-0.31}$ &$-1.75^{+0.90}_{-1.65}$  & $5.00^{+0.40}_{-0.33}$ & $1.84^{+0.14}_{-0.06}$& $-0.22^{+0.11}_{-0.14}$  & 1604.58/1940  
		\\
		223953.6-055220.0 & 4.56 & $22.58^{+0.08}_{-0.61}$ &$-0.84^{+0.12}_{-0.95}$  & $7.98^{+0.01}_{-1.48}$ & $1.83^{+0.63}_{-0.23}$& $-0.22^{+1.42}_{-0.02}$  & 483.20/1940 \\
        151002.93+570243.3&4.31 & $22.19^{+0.25}_{-0.54}$ &$-1.42^{+0.42}_{-1.10}$  & $5.96^{+0.44}_{-1.71}$ & $2.00^{+0.23}_{-0.31}$& $-0.59^{+0.41}_{-0.30}$  & 621.21/1940
		  \\
		133529.45+410125.9 & 4.26 & $22.48^{+0.01}_{-1.37}$ &$-0.93^{+0.18}_{-1.38}$  & $7.54^{+0.12}_{-1.64}$ & $1.61^{+0.62}_{-0.12}$& $0.15^{+0.28}_{-0.76}$  & 334.92/1940 \\
        132611.84+074358.3 &4.12 & $22.16^{+0.10}_{-1.26}$ &$-1.30^{+0.54}_{-1.00}$  & $6.88^{+1.04}_{-2.39}$ & $2.06^{+0.32}_{-0.21}$& $-0.34^{+0.47}_{-0.34}$  & 352.29/1940
		 \\
		163950.52+434003.7& 3.99 & $22.55^{+0.03}_{-1.47}$ &$-1.12^{+0.39}_{-0.93}$  & $6.96^{+0.95}_{-2.08}$ & $1.75^{+0.35}_{-0.06}$& $-0.09^{+0.45}_{-0.03}$  & 436.57/1940 \\
        021429.29-051744.8& 3.98 & $22.46^{+0.02}_{-1.42}$ &$-1.62^{+0.87}_{-0.60}$  & $7.36^{+0.57}_{-2.23}$ & $2.19^{+0.18}_{-0.15}$& $-0.25^{+0.47}_{-0.38}$  & 516.95/1940
		  \\
		133223.26+503431.3 & 3.81 & $22.34^{+0.19}_{-0.66}$ &$-1.48^{+0.39}_{-0.26}$  & $6.62^{+1.21}_{-0.71}$ & $2.17^{+0.19}_{-0.78}$& $-0.50^{+1.00}_{-0.50}$  & 240.41/1940 \\
        200324.1-325144.0&3.78 & $22.32^{+0.10}_{-1.04}$ &$-1.32^{+0.59}_{-0.78}$  & $6.90^{+1.04}_{-1.84}$ & $1.94^{+0.06}_{-0.33}$& $-0.08^{+0.36}_{-0.07}$  & 757.08/1940  
		\\
		200324.1-135245.1 & 3.77 & $22.25^{+0.18}_{-0.88}$ &$-1.03^{+0.30}_{-1.19}$  & $7.02^{+0.90}_{-1.30}$ & $1.80^{+0.06}_{-0.18}$& $0.05^{+0.19}_{-0.08}$  & 735.85/1940 \\
        122135.6+280614.0 & 3.31 & $22.45^{+0.04}_{-1.15}$ &$-1.15^{+0.43}_{-0.77}$  & $7.95^{+0.02}_{-2.70}$ & $1.26^{+0.21}_{-0.07}$& $0.18^{+0.16}_{-0.1 8}$  & 843.48/1940
		  \\
		042214.8-384453.0 & 3.11 & $22.31^{+0.13}_{-0.06}$ &$-1.09^{+0.37}_{-0.63}$  & $7.00^{+0.46}_{-1.20}$ & $2.17^{+0.20}_{-0.32}$& $-0.12^{+0.37}_{-0.32}$  & 527.00/1940 \\
        083910.89+200207.3 & 3.03 & $22.05^{+0.34}_{-1.27}$ &$-1.00^{+0.28}_{-1.00}$  & $6.40^{+1.45}_{-1.92}$ & $1.60^{+0.03}_{-0.38}$& $-0.32^{+0.39}_{-0.06}$  & 1048.72/1940
        \\
        111038.64+483115.6 & 2.96 & $22.32^{+0.09}_{-1.11}$ &$-1.31^{+0.59}_{-0.61}$  & $6.65^{+1.18}_{-0.69}$ & $2.50^{+0.11}_{-0.61}$& $-0.45^{+0.70}_{-0.27}$  & 333.61/1940 \\
        122307.52+103448.2 &2.75 & $22.22^{+0.05}_{-0.75}$ &$-1.34^{+0.58}_{-0.89}$  & $6.28^{+1.61}_{-0.68}$ & $2.69^{+0.18}_{-0.90}$& $-0.79^{+1.27}_{-0.07}$  & 261.18/1940
		  \\
		115005.36+013850.7 & 2.33 & $22.15^{+0.17}_{-0.45}$ &$-1.04^{+0.29}_{-0.82}$  & $6.76^{+1.10}_{-1.65}$ & $3.58^{+0.39}_{-0.50}$& $-2.02^{+1.07}_{-0.65}$  & 316.75/1940 \\
        121423.02+024252.8 &2.22 & $21.98^{+0.37}_{-0.58}$ &$-2.70^{+1.13}_{-0.30}$  & $4.98^{+0.29}_{-0.13}$ & $2.09^{+0.18}_{-0.14}$& $-0.48^{+0.21}_{-0.28}$  & 1256.46/1940  
		\\
		112338.14+052038.5 & 2.18 & $22.27^{+0.08}_{-0.91}$ &$-1.30^{+0.38}_{-0.70}$  & $6.96^{+0.70}_{-1.09}$ & $2.09^{+0.46}_{-0.38}$& $-0.27^{+0.48}_{-0.52}$  & 415.64/1940 \\
        123527.36+392824.0& 2.16 & $22.07^{+0.05}_{-1.22}$ &$-1.13^{+0.40}_{-0.76}$  & $5.38^{+0.94}_{-1.16}$ & $2.67^{+0.21}_{-0.69}$& $-0.91^{+1.03}_{-0.24}$  & 270.33/1940
		  \\
		134740.99+581242.2 & 2.05 & $22.16^{+0.14}_{-0.98}$ &$-1.19^{+0.41}_{-0.73}$  & $6.63^{+1.17}_{-0.38}$ & $2.17^{+0.26}_{-0.21}$& $-0.23^{+0.16}_{-0.38}$  & 638.71/1940 \\
        095834.04+024427.1 &1.89 & $22.38^{+0.04}_{-1.21}$ &$-1.54^{+0.76}_{-0.51}$  & $7.11^{+0.73}_{-1.54}$ & $2.27^{+0.33}_{-0.10}$& $-0.52^{+0.22}_{-0.47}$  & 500.15/1940
		 \\
		093359.34+551550.7& 1.86 & $22.04^{+0.17}_{-0.53}$ &$-1.24^{+0.52}_{-0.43}$  & $5.30^{+0.90}_{-0.45}$ & $1.29^{+0.22}_{-0.07}$& $-1.00^{+0.00}_{-0.25}$  & 659.00
		/1940 \\
        133526.73+405957.5& 1.77 & $22.45^{+0.08}_{-1.34}$ &$-1.17^{+0.45}_{-0.79}$  & $7.84^{+0.10}_{-1.42}$ & $1.58^{+0.22}_{-0.12}$& $0.11^{+0.33}_{-0.28}$  & 303.33/1940
		  \\
		100434.91+411242.8 & 1.74 & $22.46^{+0.10}_{-0.91}$ &$-0.80^{+0.07}_{-1.60}$  & $7.88^{+0.10}_{-1.42}$ & $1.81^{+0.06}_{-0.09}$& $-0.07^{+0.21}_{-0.07}$  & 1002.61/1940 \\
        104039.54+061521.5&1.58 & $22.31^{+0.02}_{-1.36}$ &$-1.30^{+0.44}_{-0.40}$  & $7.18^{+0.66}_{-1.62}$ & $2.06^{+0.42}_{-0.12}$& $0.21^{+0.33}_{-0.73}$  & 360.88/1940  
		\\
		083205.95+524359.3& 1.57 & $22.28^{+0.16}_{-0.95}$ &$-1.44^{+0.70}_{-1.56}$  & $6.89^{+0.95}_{-1.37}$ & $2.39^{+0.33}_{-0.26}$& $-0.28^{+0.49}_{-0.53}$  & 444.36/1940 \\
        112320.73+013747.4& 1.47 & $22.13^{+0.02}_{-1.02}$ &$-1.23^{+0.02}_{-0.73}$  & $6.52^{+0.24}_{-0.70}$ & $1.08^{+0.24}_{-0.14}$& $1.12^{+0.51}_{-0.03}$  & 639.3/1940
		  \\
		 091301.03+525928.9& 1.38 & $21.93^{+0.47}_{-1.93}$ &$-0.90^{+0.18}_{-1.80}$  & $7.10^{+0.82}_{-1.55}$ & $2.02^{+0.09}_{-0.07}$& $-0.44^{+0.15}_{-0.17}$  & 595.67/1940 \\
        121426.52+140258.9& 1.28 & $22.41^{+0.10}_{-1.11}$ &$-0.90^{+0.18}_{-1.25}$  & $7.79^{+0.16}_{-0.90}$ & $1.89^{+0.06}_{-0.06}$& $-0.21^{+0.17}_{-0.09}$  & 805.58/1940
        \\
        105316.75+573550.8& 1.21 & $22.06^{+0.25}_{-1.06}$ &$-1.20^{+0.44}_{-1.20}$  & $7.00^{+0.94}_{-1.13}$ & $2.19^{+0.06}_{-0.12}$& $-0.32^{+0.22}_{-0.20}$  & 588.84/1940 \\
        085808.91+274522.7&1.09 & $21.52^{+0.84}_{-0.44}$ &$-2.30^{+1.51}_{-0.70}$  & $5.06^{+2.87}_{-0.04}$ & $2.39^{+0.05}_{-0.17}$& $-0.33^{+0.34}_{-0.05}$  & 655.53/1940  
		\\
		095857.34+021314.5& 1.02 & $21.04^{+1..34}_{-0.44}$ &$-1.07^{+0.33}_{-1.07}$  & $5.58^{+2.35}_{-0.34}$ & $2.04^{+0.13}_{-0.20}$& $-0.19^{+0.29}_{-0.19}$  & 553.17/1940 \\
        125849.83-014303.3& 0.97 & $22.08^{+0.45}_{-0.66}$ &$-1.10^{+0.35}_{-0.90}$  & $7.04^{+0.39}_{-0.20}$ & $2.29^{+0.69}_{-0.06}$& $-0.23^{+0.10}_{-0.65}$  & 904.00/1940
		  \\
		 082257.55+404149.7& 0.87 & $21.32^{+0.86}_{-0.15}$ &$-0.91^{+0.17}_{-1.49}$  & $5.04^{+2.84}_{-0.0}$ & $2.51^{+0.07}_{-0.45}$& $-0.90^{+0.72}_{-0.05}$  & 365.24/1940 \\
         150431.30+474151.2&0.82 & $21.71^{+0.74}_{-0.53}$ &$-1.62^{+0.90}_{-0.48}$  & $5.59^{+2.37}_{-0.00}$ & $2.15^{+0.73}_{-0.19}$& $-0.11^{+0.44}_{-0.66}$  & 407.23/1940
		 \\
		111606.97+423645.4& 0.67 & $21.66^{+0.51}_{-0.96}$ &$-1.13^{+0.36}_{-0.83}$  & $5.84^{+1.90}_{-0.22}$ & $1.94^{+0.06}_{-0.23}$& $-0.26^{+0.28}_{-0.08}$  & 680.82
		/1940 \\
        130028.53+283010.1& 0.65 & $21.36^{+0.97}_{-0.89}$ &$-1.00^{+0.27}_{-1.99}$  & $6.88^{+1.07}_{-0.82}$ & $1.96^{+0.04}_{-0.04}$& $-0.06^{+0.08}_{-0.10}$  & 971.87/1940
		  \\
		111135.76+482945.3 & 0.56 & $20.85^{+1.35}_{-0.54}$ &$-0.99^{+0.23}_{-1.17}$  & $7.83^{+0.04}_{-2.44}$ & $2.26^{+0.11}_{-0.04}$& $-0.12^{+0.13}_{-0.17}$  & 633.63/1940 \\
        091029.03+542719.0 &0.53  & $21.34^{+0.98}_{-0.30}$ &$-1.72^{+0.96}_{-0.68}$  & $5.07^{+2.80}_{-0.03}$ & $2.61^{+0.08}_{-0.25}$& $-0.68^{+0.55}_{-0.04}$  & 599.13/1940  
		\\
		105224.94+441505.2& 0.44 & $21.18^{+1.09}_{-0.48}$ &$-0.88^{+0.16}_{-1.08}$  & $5.99^{+1.88}_{-0.49}$ & $2.48^{+0.14}_{-0.19}$& $-0.40^{+0.44}_{-0.26}$  & 419.56/1940 \\
        223607.68+134355.3& 0.33 & $21.86^{+0.37}_{-1.87}$ &$-0.84^{+0.12}_{-1.21}$  & $6.89^{+0.95}_{-1.84}$ & $2.69^{+0.19}_{-0.03}$& $-0.20^{+0.12}_{-0.45}$  & 510.23/1940
		  \\
		144645.93+403505.7& 0.27 & $20.60^{+0.98}_{-0.90}$ &$-1.85^{+0.30}_{-1.85}$  & $5.03^{+0.09}_{-0.01}$ & $2.98^{+0.02}_{-0.10}$& $-0.71^{+0.19}_{-0.04}$  & 885.25/1940 \\
        123054.11+110011.2& 0.24 & $21.15^{+0.73}_{-0.85}$ &$-2.52^{+1.03}_{-0.48}$  & $5.07^{+0.13}_{-0.04}$ & $2.57^{+0.17}_{-0.07}$& $-0.59^{+0.14}_{-0.07}$  & 679.62/1940
        \\
        103059.09+310255.8& 0.18 & $20.90^{+1.06}_{-0.43}$ &$-2.80^{+0.82}_{-0.16}$  & $5.06^{+0.14}_{-0.01}$ & $2.28^{+0.05}_{-0.06}$& $-0.53^{+0.08}_{-0.06}$  & 1514.88/1940 \\
        141700.81+445606.3& 0.11 & $21.00^{+0.66}_{-1.00}$ &$-2.85^{+0.87}_{-0.15}$  & $5.09^{+0.06}_{-0.03}$ & $2.70^{+0.04}_{-0.07}$& $-0.57^{+0.10}_{-0.07}$  & 1216.11/1940
        \\
        
	\hline
	\end{tabular}
    
\end{table*}

In this section, we discuss the result of using a log-parabolic power law compared to the more commonly used simple power law for the QSO intrinsic  continuum in Section \ref{subsec:4.1}. We give the IGM property results for the full sample using the CIE absorption model in Section \ref{subsec:4.2}. All spectral fits include Galactic and QSO host CGM absorption as described in Section \ref{sec:models}.

\subsection{Spectra fits using alternative continuum models\label{subsec:4.1}}
In nearly all of the sample, the Cstat fit improved using the log-parabolic power law with $~60\%$ showing a significant improvement based on the criteria $\Delta$Cstat$ > 2.71$. Accordingly, in fitting the QSO sample with the full CIE model, we used only a log-parabolic power law for consistency.

\subsection{Results for IGM parameters using the CIE model}\label{subsec:4.2}
\begin{figure} 
    \centering
    
    \includegraphics[scale=0.55]{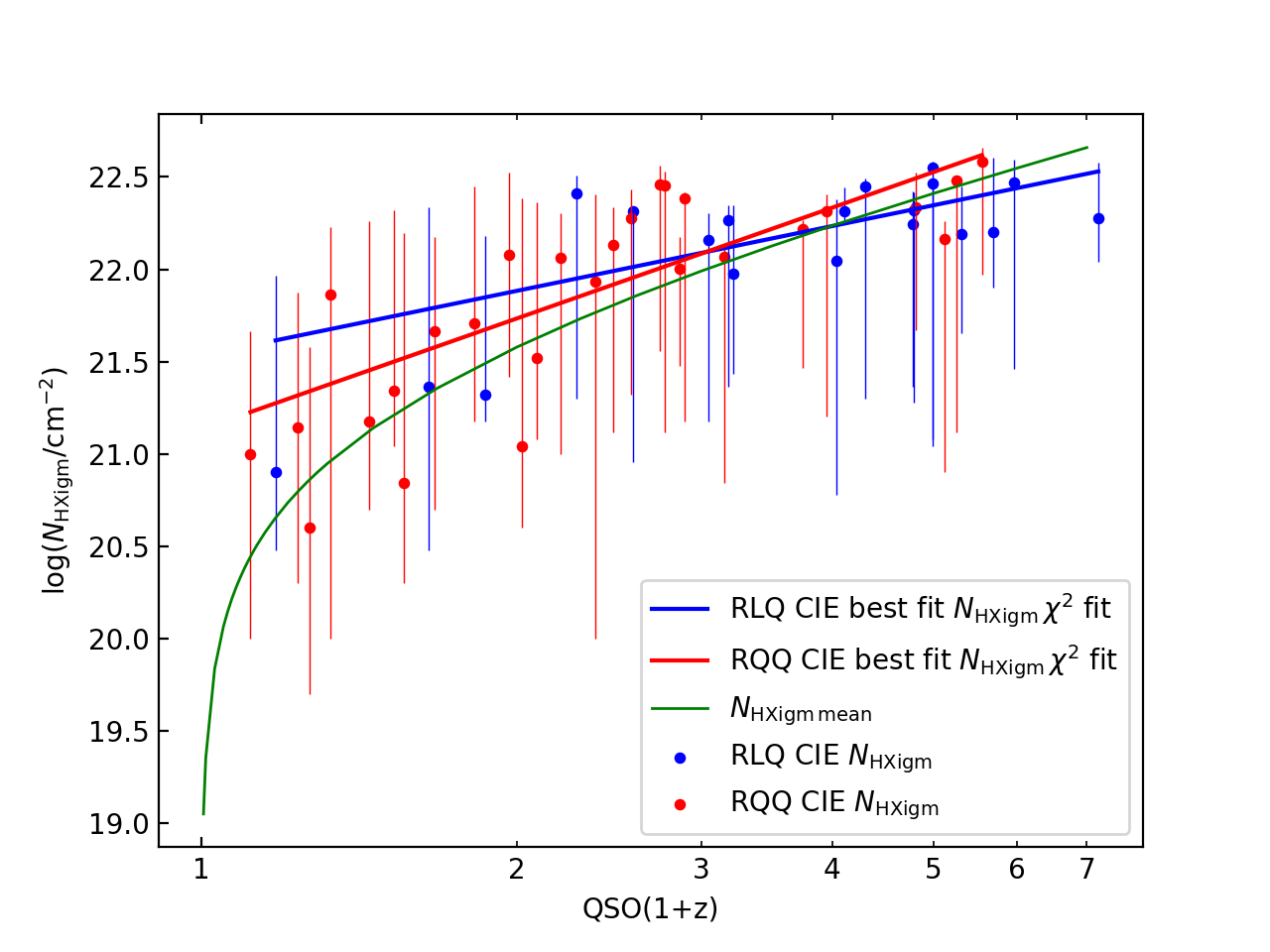} 
    \\
    
    \caption{Results for the IGM $\textit{N}\textsc{hx}$ parameter versus redshift using the CIE (\textsc{hotabs}) model. RLQ are blue and RQQ are red. The error bars are with a $90\%$ confidence interval. The green line is the simple IGM model (see eq. \ref{eq:simpleIGM}).}
        \label{fig:NHX_z}
\end{figure}

Table \ref{tab:Fullresults} gives the results for log($\textit{N}\textsc{hx}$), temperature and metallicity using the CIE IGM model component for our full QSO sample. These IGM parameters, as well as the power law parameters were left free to vary. The error bars are with a $90\%$ confidence interval. The green line in the plots of $\textit{N}\textsc{hx}$ and redshift (Figs. \ref{fig:NHX_z} and \ref{fig:QSO_FSRQ_GRB_NHX_z}), is the simple model of the mean IGM hydrogen density (eq.\ref{eq:simpleIGM}) based on D20, D21a and D21b and references therein \citep[e.g.][]{Starling2013,Shull2018}.
\begin{equation} 
\label{eq:simpleIGM}
\mathit{N}_{\textsc{hxigm}} = \frac{n_0 c}{H_0} \int_0^z \frac{(1 + z)^2 dz}{[\Omega _M(1 + z)^3 + \Omega _\Lambda ]^\frac{1}{2}}
\end{equation}
where n$_{0}$ is the hydrogen density at $z = 0$, taken as $1.7 \times 10^{-7}$ cm$^{-3}$ \citep{Behar2011}. We used our results for $\mathit{N}\textsc{hxigm}$ and actual redshift for the QSOs to get their equivalent $n_0$ which are are derived by rearranging equation \ref{eq:simpleIGM} to give $n_0$.  We then took the mean of $n_0$ for our full sample and calculated the standard error.

\begin{figure} 
    \centering
    
    \includegraphics[scale=0.55]{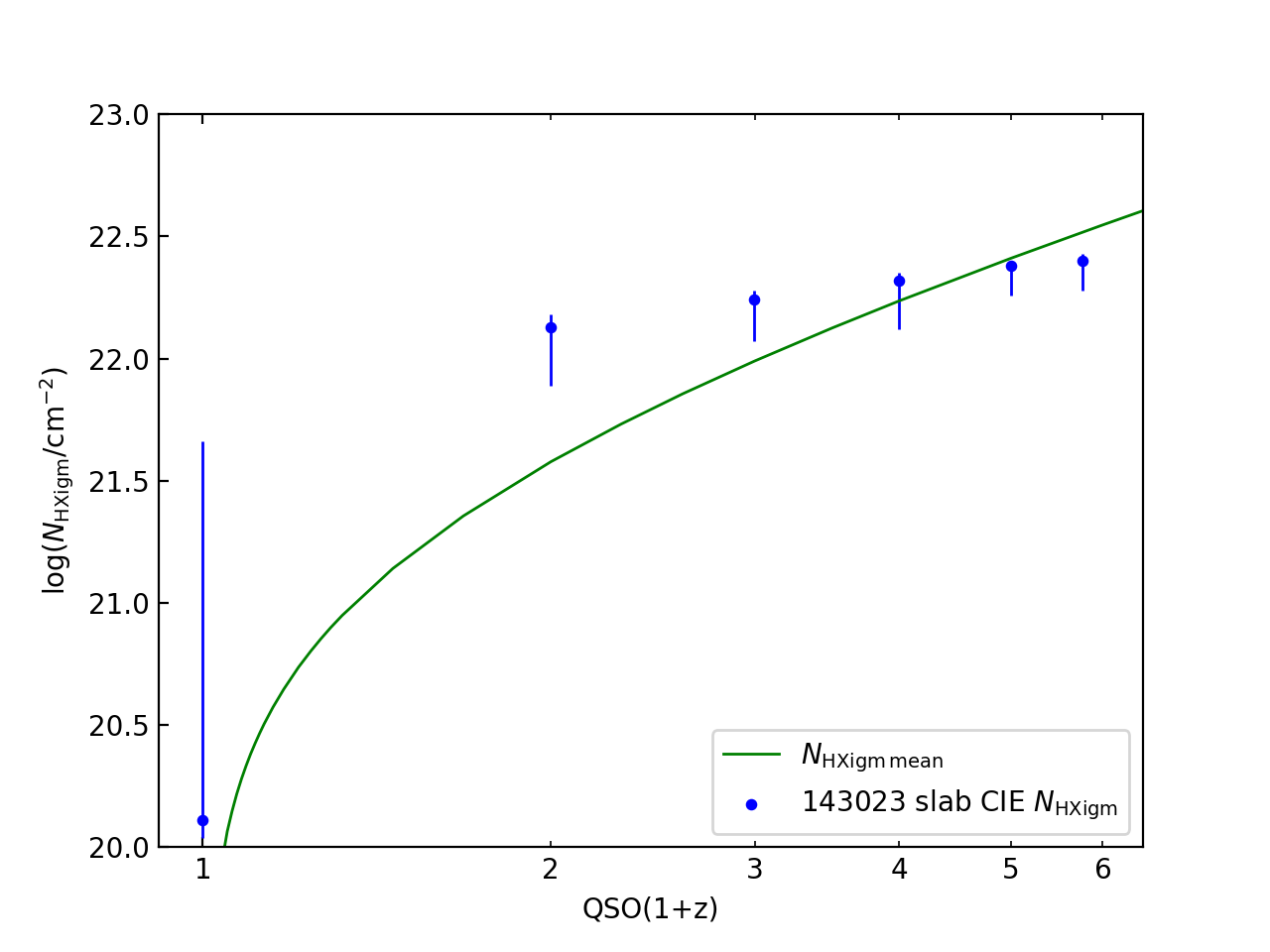} 
    \caption{The impact on $\mathit{N}\textsc{hxigm}$ for 143023.73+420436.5 by moving the IGM slab from $z = 0$ to 4.71, freezing log(T/K) = 6 and $[X/$H$] = -1$. The green line is the simple IGM model.}
        \label{fig:143023slab}
\end{figure}

In Fig. \ref{fig:NHX_z}, the $\mathit{N}\textsc{hxigm}$ versus redshift for the full QSO sample scales as $(1 + z)^{1.5\pm0.2}$, reduced $\chi^2 = 0.58$ (approximated $\chi^2$ given the uncertainties are uneven). For the RQQ dominating at redshift $z < 2$, the redshift scaling is $(1 + z)^{1.9\pm0.3}$, while the RLQ dominating at $z > 2$ scale as $(1 + z)^{1.2\pm0.3}$.  This scaling of $\mathit{N}\textsc{hxigm}$ is very similar to the simple IGM model curve (reduced $\chi^2 = 1.78$), subject to error bars i.e. it is what is expected for a diffuse IGM.  The sample includes QSOs with redshift to $z = 0.114$, so a linear $\chi^2$ fit is only an approximation for the curve. The mean hydrogen density based on eq.\ref{eq:simpleIGM} for the QSO sample is $n_0 = (2.8\pm{0.3}) \times 10^{-7}$ cm$^{-3}$ at $z = 0$, compared to $1.7 \times 10^{-7}$ cm$^{-3}$ assumed for the simple  IGM model. A sub-sample of QSOs with $z >1.6$, similar to the GRB sample in D21a, gives $n_0 = (2.1\pm0.3) \times 10^{-7}$ cm$^{-3}$. 

Most X-ray absorption occurs below 2 keV in the rest frame. Given we are using observed 0.3-10 keV spectra, for higher redshift QSOs, the slab location assumption results in lower keV absorbing ions being redshifted out of the observed spectral range. Placing the slab at less than half the QSO redshift may better trace the low keV X-ray absorption. However, it would not reflect the impact on the observed cross-section which scales approximately as $E^{-2.5}$, and therefore for redshifted absorbers with a fixed observed energy window, the cross section scales as $\sim(1 + z)^{-2.5}$. To show the impact of placing the slab at different redshifts, other than the model location of half the QSO redshift, we used QSO 143023.73+420436.5 which is located at $z = 4.71$ as an example. We fitted the spectrum moving the IGM slab from $z = 0$ to 4.71, freezing log(T/K) = 6 and $[X/$H$] = -1$. As can be seen, the $\mathit{N}\textsc{hxigm}$ is not substantially affected by the choice of redshift location, apart from at $z = 0$ which would not reflect any IGM absorption. The uncertainties are smaller as there are less free parameters than the full free model.

 \begin{figure} 
    \centering
    
    \includegraphics[scale=0.55]{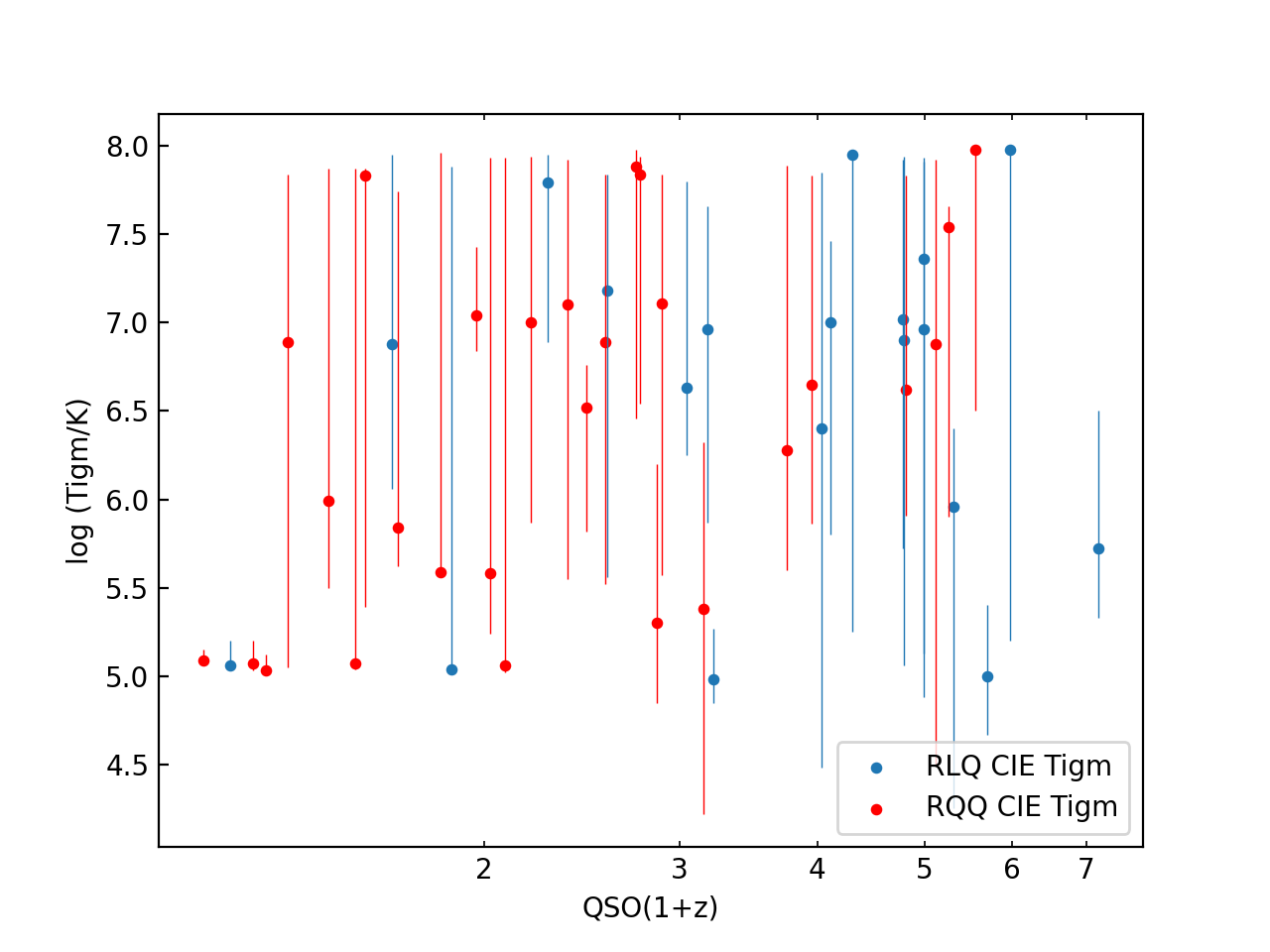} 
    \caption{Results for the log(T/K) IGM versus redshift using the CIE model. RLQ are blue and RQQ are red. The error bars are with a $90\%$ confidence interval. The fit was too poor for a $\chi^2$ curve due to the large scatter.}
        \label{fig:IGM_Tandz}
\end{figure}

\begin{figure} 
    \centering
    
    \includegraphics[scale=0.55]{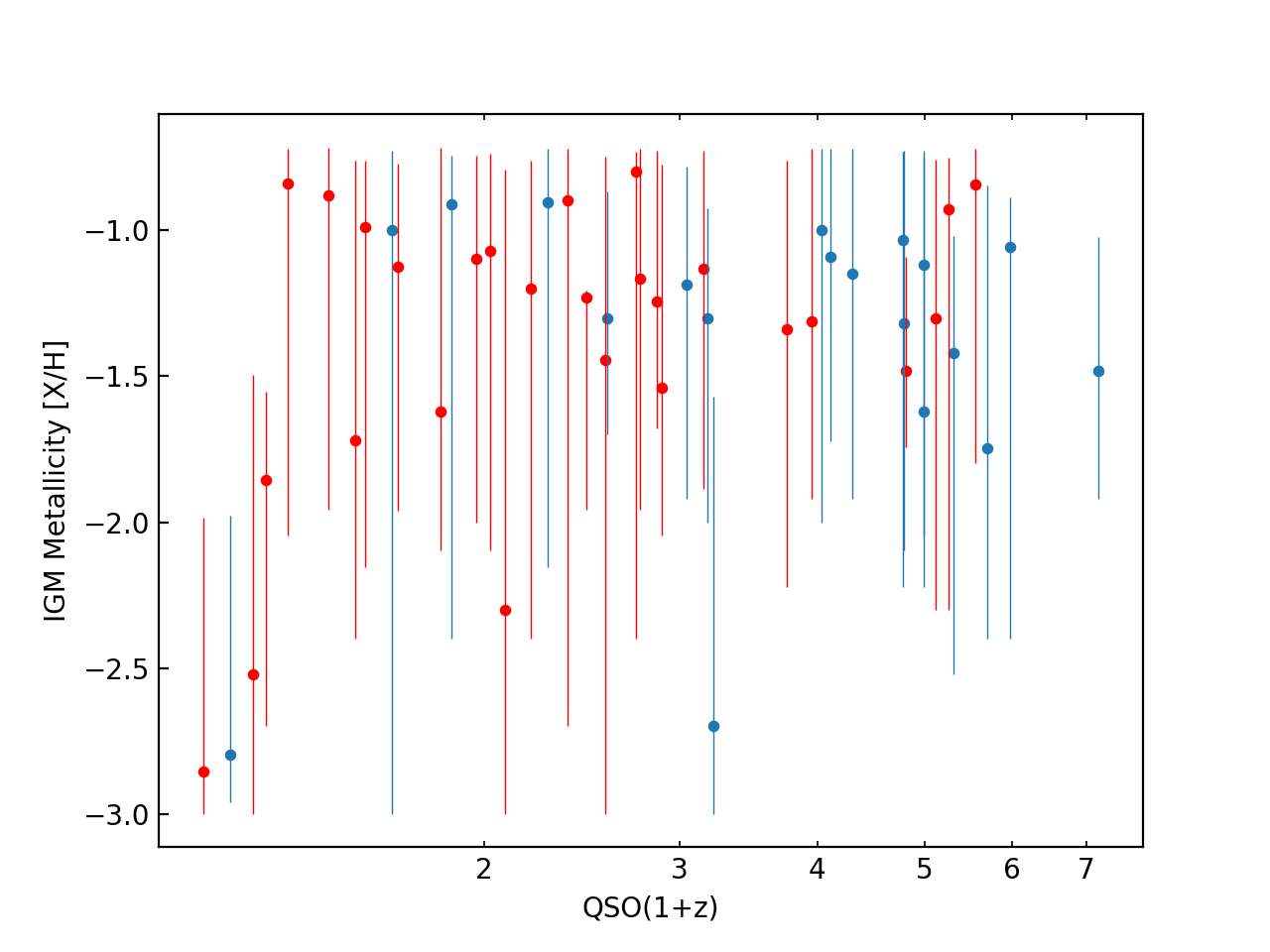} 
    \caption{IGM metallicity versus redshift using the CIE model. RLQ are blue and RQQ are red. The error bars are with a $90\%$ confidence interval. We do not include a $\chi^2$ curve in the plot as the fit was poor due to the large scatter.}
        \label{fig:IGM_Zandz}
\end{figure}

 There is a broad range in the temperature across the redshift range for the QSO sample $4.9 <$ log($T$/K) $< 8.0$, with most having large error bars in Fig. \ref{fig:IGM_Tandz}. The mean temperature for the full QSO sample is log($T$/K) $= 6.5\pm{0.1}$. It is notable that very few of the QSOs have error bars that go below log($T$/K) $< 5.0$ even though we allow the temperature parameter to vary down to log($T$/K) $= 4.0$. A number of the QSOs have best fit temperatures close to the high or low parameter range limits, indicating that temperatures are not well determined.
 
 No relation between temperature and redshift is apparent.  The IGM LOS may include a cooler photo-ionised  gas contributing to the absorption which is not included in this CIE model. The fits are not representative of any individual absorber temperature, but instead represent the integrated LOS.

 \begin{figure*} 
    \centering
    \begin{tabular}{c|c}
    \includegraphics[scale=0.55]{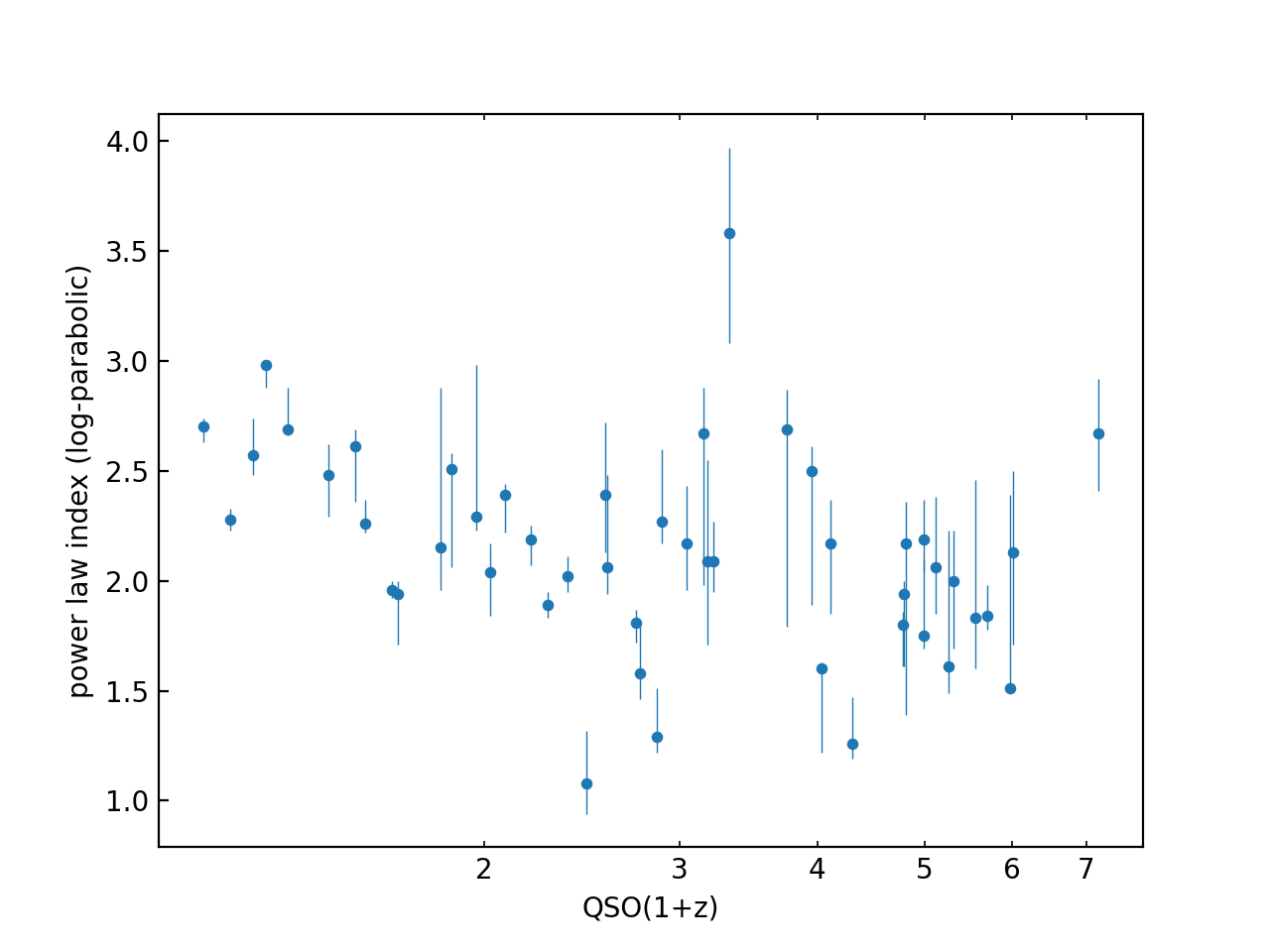} 
    \includegraphics[scale=0.55]{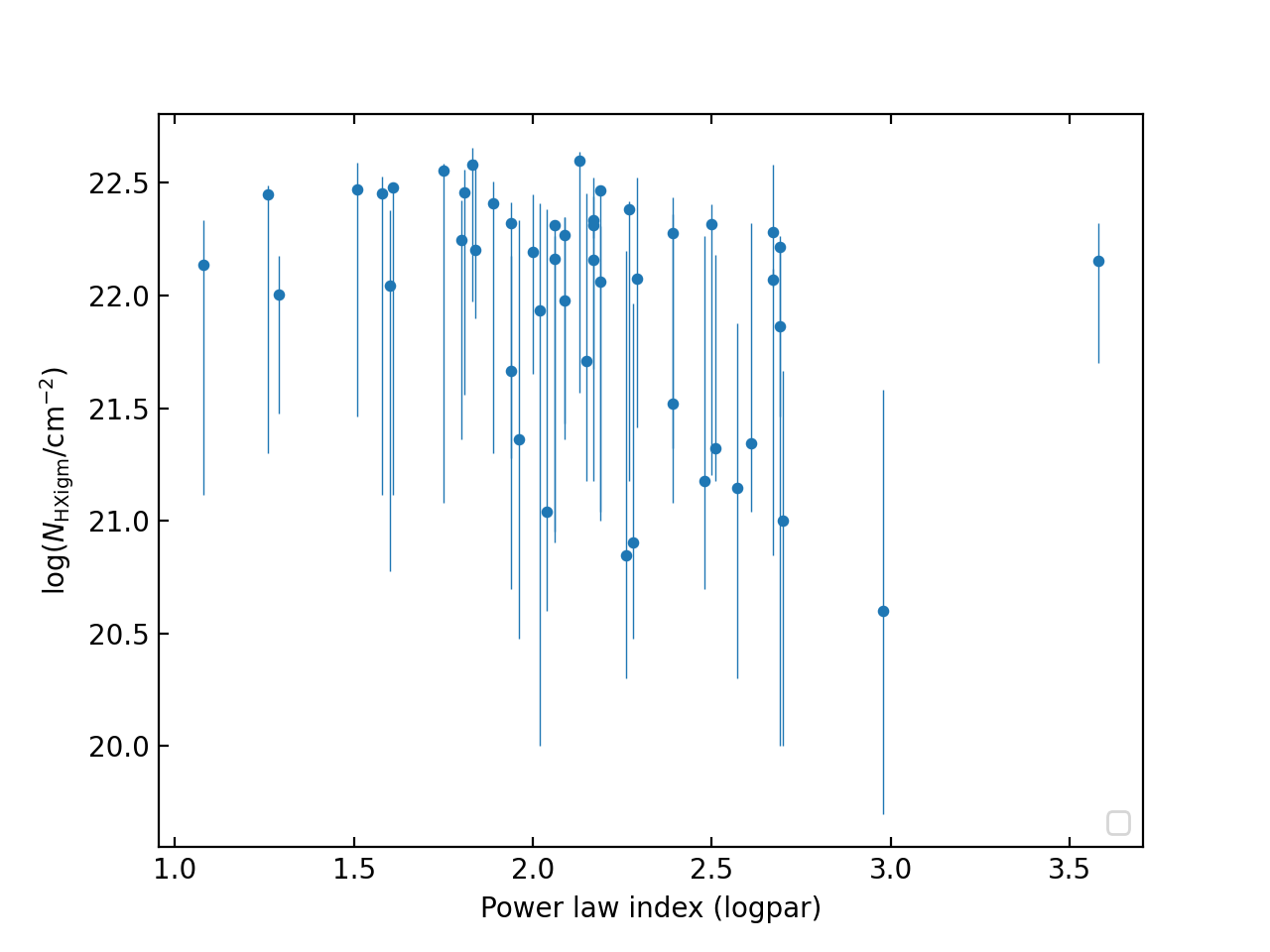}
    \end{tabular}
    \caption{Testing for a possible relation between $\mathit{N}\textsc{hxigm}$ and the QSO intrinsic power law index (log-parabolic). Left panel is the QSO intrinsic power law index versus redshift. There is no apparent strong relation between the power law and redshift, other than that due to the dominance of RQQ below $z <2$ which are known to have a higher power law index than RLQ. The right panel is $\mathit{N}\textsc{hxigm}$ and the QSO intrinsic power law index which does not show any apparent relation between the variables. }
        \label{fig:logPO_NHx}
\end{figure*}

 There is no apparent relation between $[X/$H$]$ and redshift in Fig. \ref{fig:IGM_Zandz}. The mean metallicity is $[X/$H$] = -1.31\pm0.07$ (0.05Z\sun) and ranges from approximately $[X/$H$] -0.8\/\ (0.16Z\sun)$ to $[X/$H$] -2.9\/\ (0.001Z\sun)$. Most of the QSOs appear to favour metallicity in the range $-1 \leq [X/$H$] \leq-2.0$, with only a small number favouring lower metallicities, generally at lower redshifts. This appears to be contrary to any expected  evolution of metallicity, though our approach is based on the full LOS to the QSOs and not any particular absorber redshift.

Based on our results, the CIE model using \textsc{hotabs} is plausible for modelling the warm/hot component of the IGM at all redshifts, with the caveats of using only a CIE IGM component, the slab model being representative of the full LOS, and low X-ray resolution. In particular, we note that the results are sensitive to the assumption that placing the slab at half the QSO redshift is representative of the diffuse IGM. 

In all fits, the Cstat was improved by using the IGM component. Further, $73\%$ show a significant improvement with the IGM component added based on the criteria $\Delta$Cstat$ > 6.25$ for three interesting parameters. The average Cstat improvement for the full sample per free IGM parameter was 8.25. Our metallicity and temperature ranges, and mean results are consistent with simulations for a warm/hot phase, with the caveat that we model the continuum curvature and not specific absorption features  so there is scope for degeneracy in the three free IGM parameters. 

In Section \ref{sec:Robust} we test the robustness of our results.

\begin{figure*} 
    \centering
    \begin{tabular}{c|c}
    \includegraphics[scale=0.55]{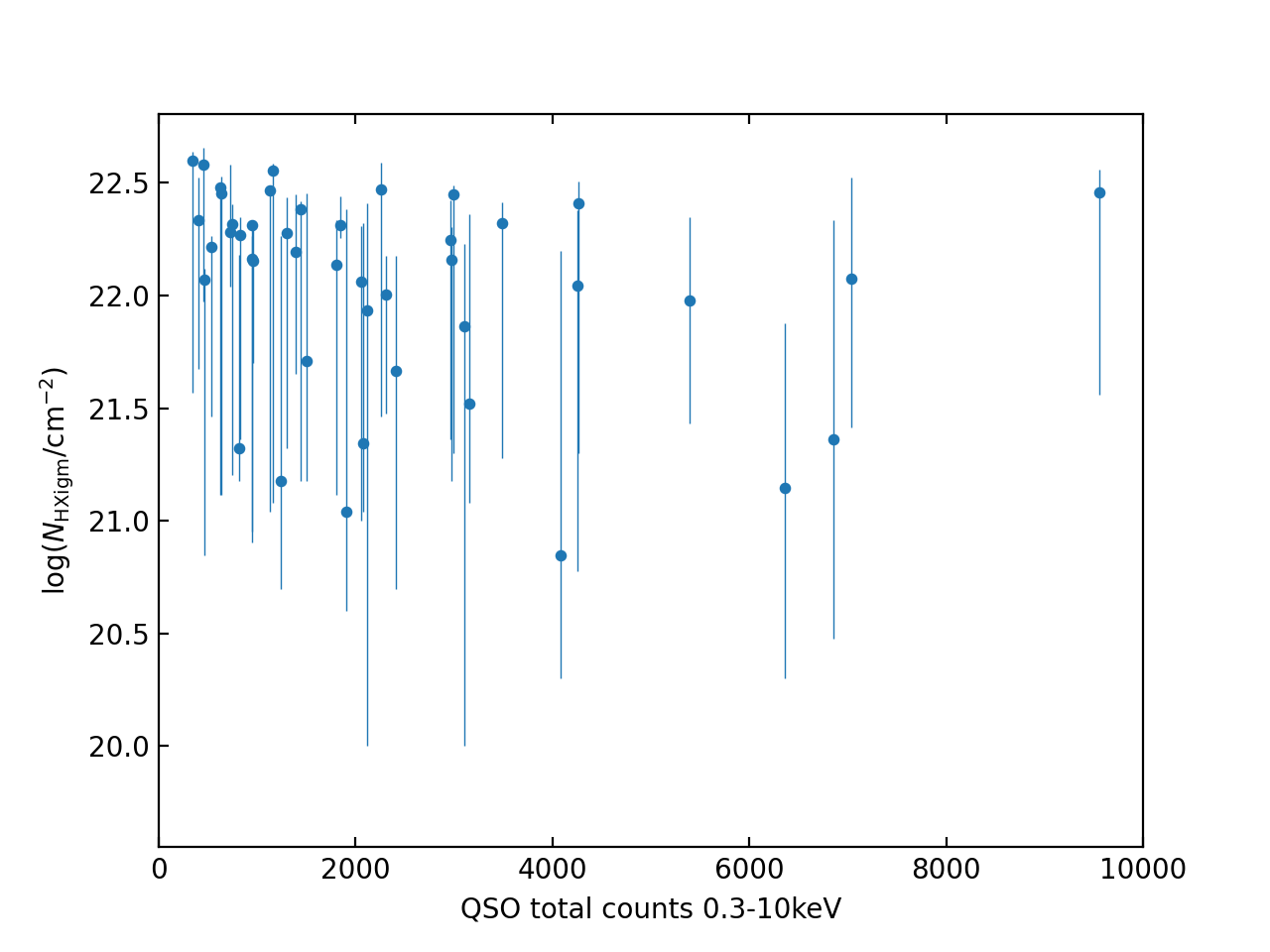}
    \includegraphics[scale=0.55]{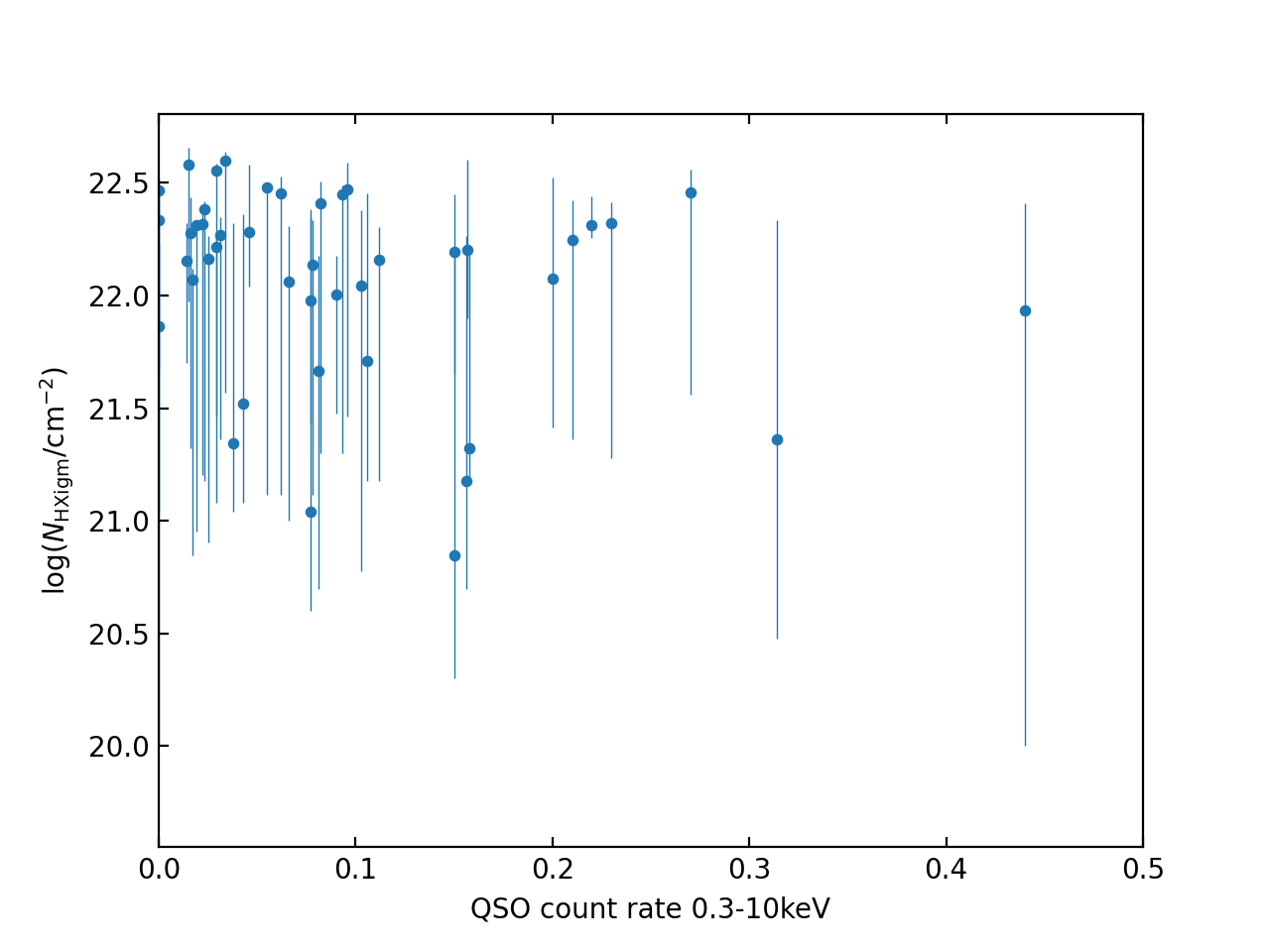}
    \end{tabular}
    \caption{Left panel - testing for any relation between the QSO $\mathit{N}\textsc{hxigm}$ and total spectral counts. Right panel - testing for any relation between $\mathit{N}\textsc{hxigm}$ and spectral count rates.  There is no apparent relation between the variables.}
        \label{fig:NHXigm_counts}
\end{figure*}

\section{Tests for robustness of IGM parameter results}\label{sec:Robust}

AGN are generally known to have a Compton hump at higher energies. Similarly, at lower energies, a soft excess is sometimes observable in AGN spectra, whose cause is still debated. Both, or either, of these features, if present in a QSO spectra, may affect any absorption feature of the IGM. $\mathit{N}\textsc{hxigm}$ may be degenerate with continuum slope and intrinsic curvature. Further, QSO spectra have very large differences in total counts and count rates which could have an impact on or be linked to spectral curvature. There is a large range in luminosity of QSOs and this may be a source of the apparent $\mathit{N}\textsc{hxigm}$ redshift relation. Finally, we look for absorbers in UV and lensing galaxies to investigate their possible contribution to the column density.

\begin{figure*} 
    \centering
    \begin{tabular}{c|c}
    \includegraphics[scale=0.55]{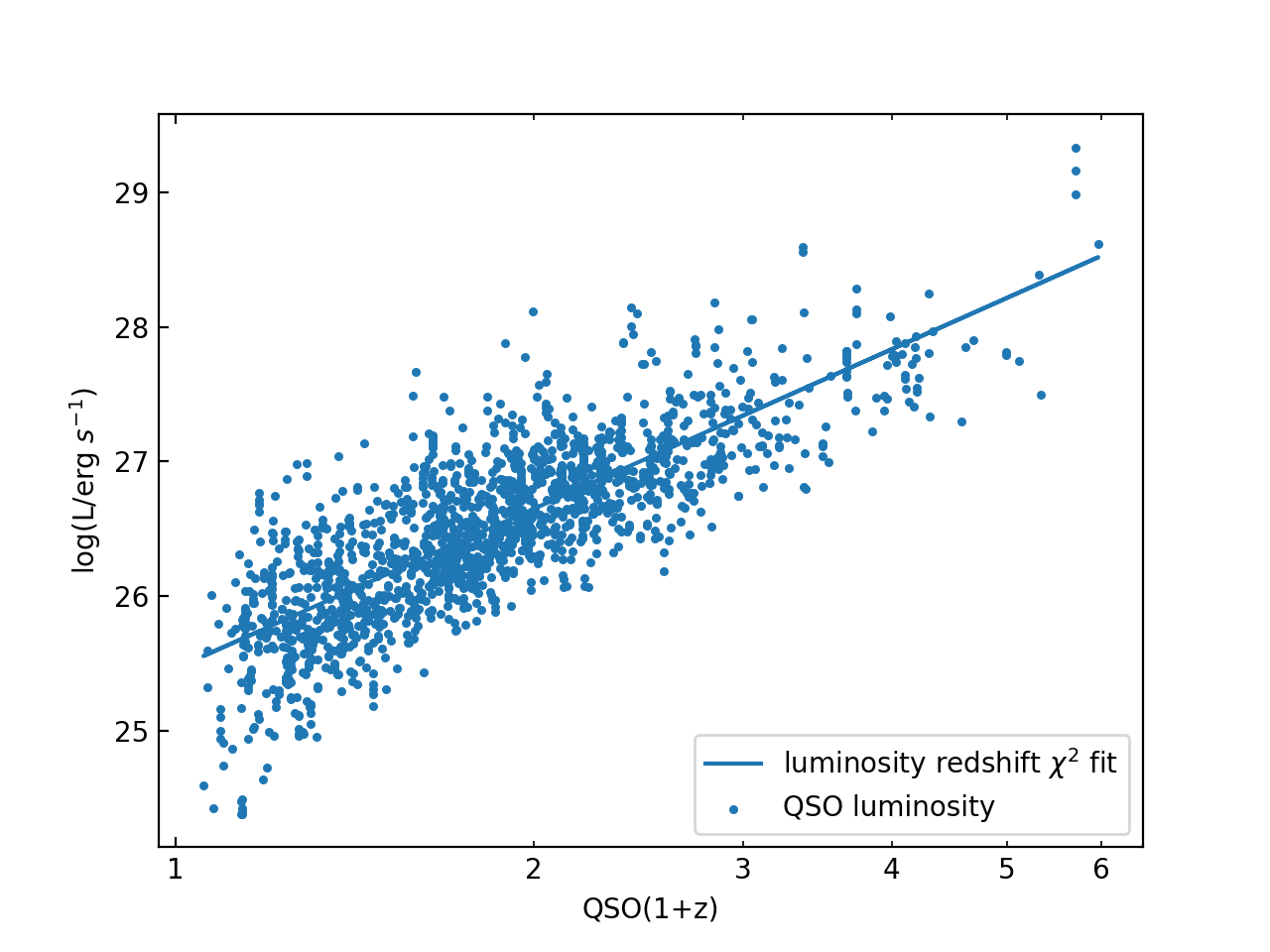} 
    \includegraphics[scale=0.55]{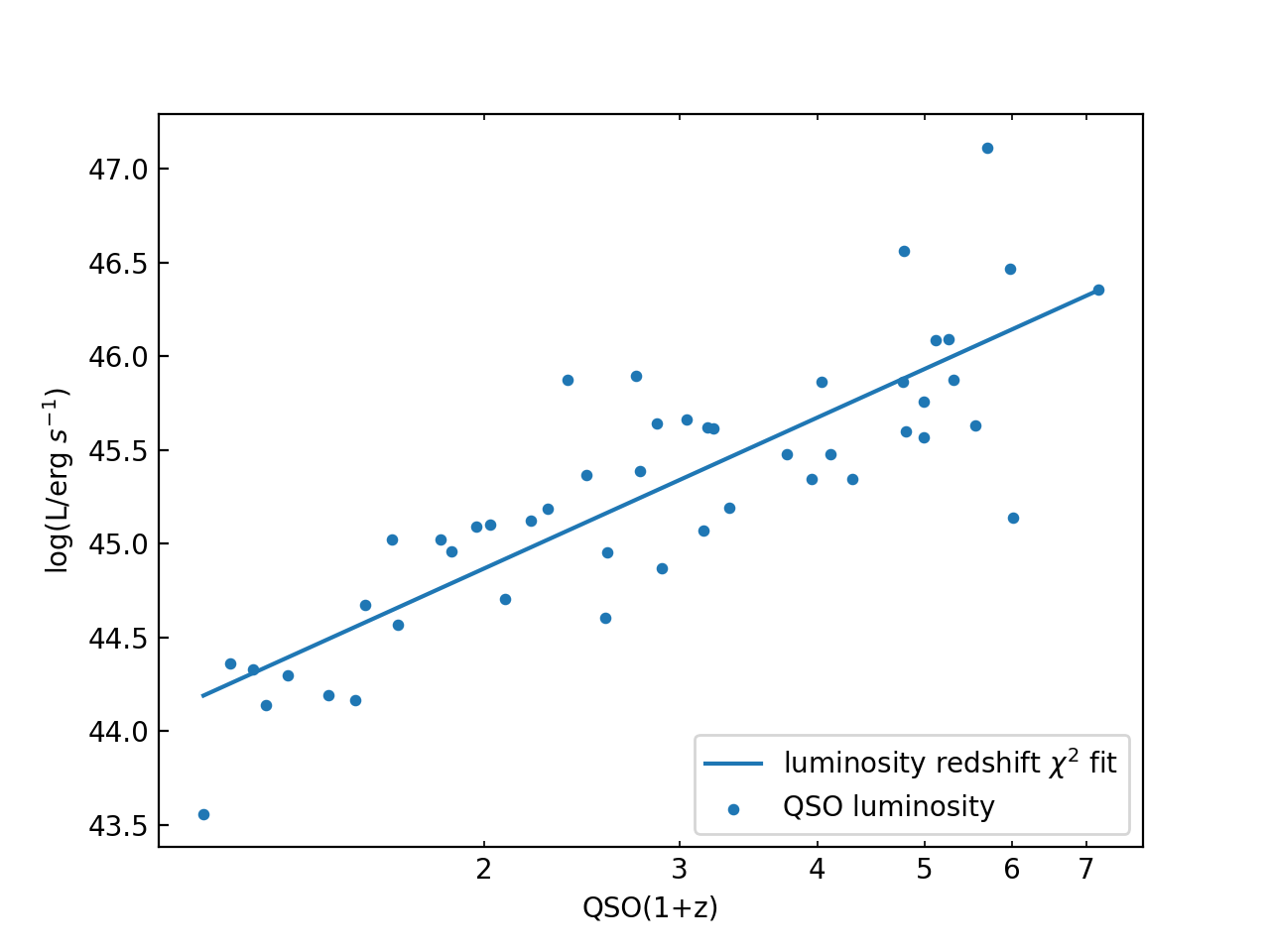}
    \end{tabular}
    \caption{Left panel - Monochromatic luminosity (2keV) and redshift for the full SDSS-DR14 and $\textit{ 4XMM-DR9}$ with counts > 1000. Right panel - Our QSO sample luminosity for energy 2 - 10keV and redshift. }
        \label{fig:L_z}
\end{figure*}

\begin{figure} 
    \centering
    
    \includegraphics[scale=0.55]{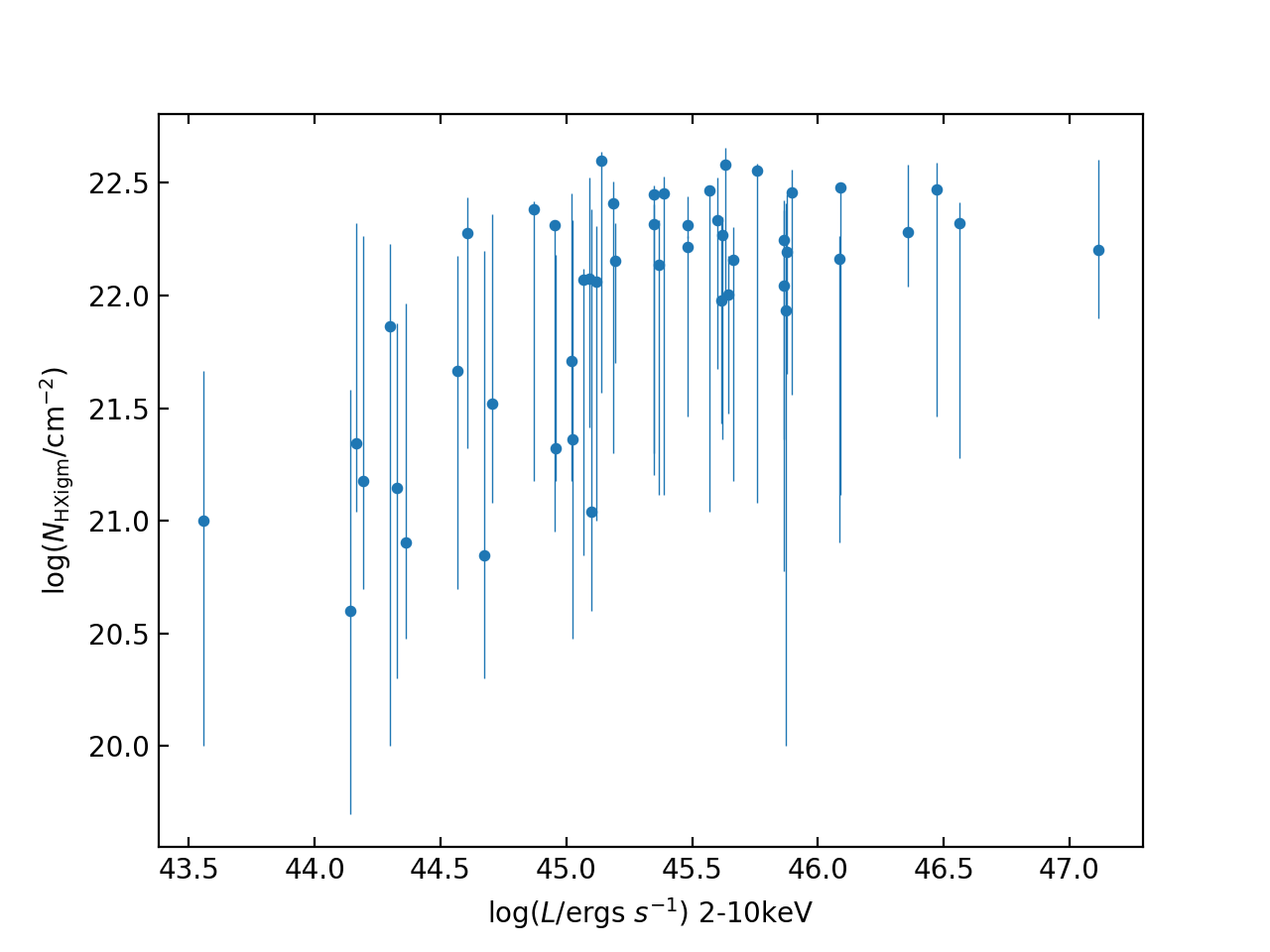} 
    \caption{Testing for a possible relation between $\mathit{N}\textsc{hxigm}$ and luminosity.}
        \label{fig:NHXigm_lum}
\end{figure}

\subsection{IGM column density and intrinsic power law index}

There is scope for degeneracy between $\mathit{N}\textsc{hxigm}$ and spectral slope and curvature. To measure the unabsorbed continuum slope, we used a log-parabolic power law only. RQQ are dominant at $z < 2$, and typically have a higher power index than RLQ which dominate above above $z > 2$ \citep[e.g.][]{Scott2011a,Page2005,Reeves2000}, which is consistent with our results. Fig \ref{fig:logPO_NHx} left panel does not show a strong relation between the QSO power law index and redshift, other than that expected from the redshift spread of RQQ and RLQ.  There have been many studies over the years examining possible evolution or relations of QSO continua with redshift. The results consistently have been that there is no such evolution or relation  \citep[e.g.][]{Reeves2000,Page2005,Grupe2006,Risaliti2019,Shehata2021c}. This supports the argument that the observed $\mathit{N}\textsc{hx}$ redshift relation in our results is IGM related and not intrinsic to the QSO properties, as there is no apparent relation between our $\mathit{N}\textsc{hxigm}$ results and power law index in Fig. \ref{fig:logPO_NHx} right panel.

\subsection{IGM column density and spectral counts}
Since the QSO spectra and therefore $\mathit{N}\textsc{hxigm}$ will be better constrained for observations with higher statistics, total counts and count rates, this may lead to a bias. We check this by looking for any relation between the index and net counts and count rates. Fig \ref{fig:NHXigm_counts} left panel shows no relation between the QSO $\mathit{N}\textsc{hxigm}$ and total counts. There is also no apparent relation between $\mathit{N}\textsc{hxigm}$ and count rates in Fig. \ref{fig:NHXigm_counts} right panel, as expected given that there is no obvious physical reason why a higher flux should be linked to column density, consistent with prior studies \citep[e.g.][]{Shehata2021c}.

\subsection{IGM column density and luminosity}
The majority of QSOs are RQQ with only approximately $10\%$ being RLQ, though this varies somewhat with redshift \citep[e.g.][]{Grupe2006}. However, RQQ are mostly observed at $z < 2$. Given we are selecting the QSOs with the highest counts, and also out to the highest redshifts, there is scope for luminosity bias which may be degenerate with $\mathit{N}\textsc{hxigm}$. In Fig. \ref{fig:L_z} left panel, we plot the SDSS-DR14 and 4$\textit{XMM-Newton}$-DR9 cross-correlated catalogue with a cutoff of $>1000$ counts. A clear luminosity redshift relation is notable. In Fig. \ref{fig:L_z} right panel, a plot of our QSO sample with redshift shows a similar luminosity redshift relation.

In Fig. \ref{fig:NHXigm_lum}, we plot our QSO $\mathit{N}\textsc{hxigm}$ and luminosity. Given the luminosity bias in our sample, and the observed luminosity redshift relation in both our sample and the full SDSS-$\textit{XMM-Newton}$ catalogue, it is not surprising that there is also an apparent $\mathit{N}\textsc{hxigm}$ luminosity relation.
This relation has been noted previously and it is not possible to determine which parameter of either luminosity or redshift, that $\mathit{N}\textsc{hxigm}$ is more closely related  \citep{Eitan2013, Shehata2021c}, or whether the luminosity relationship is causal in any way on $\mathit{N}\textsc{hxigm}$. The results of $\mathit{N}\textsc{hxigm}$ redshift relations for other tracers in Section \ref{sec:combinedQSOGRBblazar} should help clarify this point. In that section, we note that GRBs, blazars and QSOs all show a very similar consistent relation between $\mathit{N}\textsc{hxigm}$ and redshift supporting the argument that the rising $\mathit{N}\textsc{hxigm}$ is not caused by luminosity.

\subsection{Compton reflection hump}

A Compton or reflection hump feature is common in AGNs at a rest frame of $~30$keV. Depending on the redshift, this could appear in the spectra observed frame between $0.3 - 10$keV, especially above $z > 3$. The most common model in \textsc{xspec} used for this feature is \textsc{pexrav} \citep{magdziarz1995}, which assumes an optically thick, cold material,  distributed in a slab. In our test fitting, we left the parameter R (slab scaling parameter) free, with the power law and normalisation tied to continuum power law, and with the other parameters set to \textsc{xspec} default values following the conventional approach \citep[e.g.][]{Reeves2000,Ricci2017}. We refitted all our QSO sample with \textsc{pexrav} instead of our CIE IGM component. For most of the QSOs, the Cstat fit was worse with \textsc{pexrav}. For all QSOs with $z > 3$, the reflection parameter results were R $\ll 1$. For the small number of QSOs that had similar Cstat results as for the CIE IGM model, a visual inspection of the spectra indicated a possible Fe feature at a rest frame of $6 - 7$ keV. When refitted with the CIE IGM component added, the IGM parameters did not change i.e. the inclusion of the relection component did not impact the result. For the two lowest redshift QSOs, the Cstat fit improved significantly with both \textsc{pexrav} and our CIE IGM model included (being $\Delta$Cstat = 11.5 and 7.7 for QSOs 103059.09+310255.8 and 141700.81+445606.3 respectively). However, again, for these two low redshift QSOs, the IGM fit parameters did not alter substantially.

Our results are consistent with previous studies for QSOs which found that the reflection component was very weak or consistent with no reflection in both RQQ and RLQ and that this was inversely related to luminosity, known as the X-ray Baldwin effect \citep[e.g.][]{Reeves2000,Iwasawa}.

\subsection{Soft excess}

Many AGN show a soft excess in their X-ray spectra e.g. \citet[]{Ricci2017} who found that over $50\%$ of their AGN showed evidence of soft excess. However, their AGN sample was restricted to $z <0.3$. The soft excess is typically modelled as a blackbody as a simple representation, with a peak rest frame temperature of $\sim0.1$keV, and a range of $~0.01 \leq kT \leq 0.3$keV \citep[e.g.][]{Scott2011a,Reeves2000,Ricci2017}.

We refitted all our sample with a redshifted blackbody, $\textsc{zbbody}$ in $\textsc{xspec}$ instead of the IGM component. None of our QSOs with $z >0.3$ showed any evidence of a soft excess consistent with previous studies \citep[e.g.][]{Reeves2000}. For the 4 QSOs with $z <0.3$ the inclusion of a blackbody with a simple power law, did improve the Cstat fit. For the three lowest redshift QSOs, the inclusion of the IGM component as well as a blackbody component significantly improved the fit ($\Delta$Cstat > 6.25). However, the IGM fitted parameters did not alter substantially. 

\subsection{Large absorbers on the line of sight}
To test whether DLAs could account for some of the absorption on the LOS, we reviewed the SDSS spectra\footnote{http://skyserver.sdss.org/dr16/en/tools/explore/} for evidence of DLAs and cross-checked with the new catalogue based on SDSS DR16Q \citep{Ho2021}. Our results for DLAs on the QSO LOS were consistent with \citep{Prochaska2018}, who were investigating the average number of DLAs intersected by a LOS to a source at redshift out to $z \sim 5$. We found 1 QSO with a DLA between $2 \leq z \leq 3$, and 8 QSOs with DLAs between $3 \leq z \leq 5$. None had more than 2 DLAs on a particular QSO LOS. All of these QSOs in our sample, which showed DLAs from the SDSS, had log($\mathit{N}\textsc{hxigm}$/cm$^{-2}) > 22$, and therefore the DLA contribution would be insignificant to the column.

\begin{figure*} 
    \centering
    \begin{tabular}{c|c}
    \includegraphics[scale=0.55]{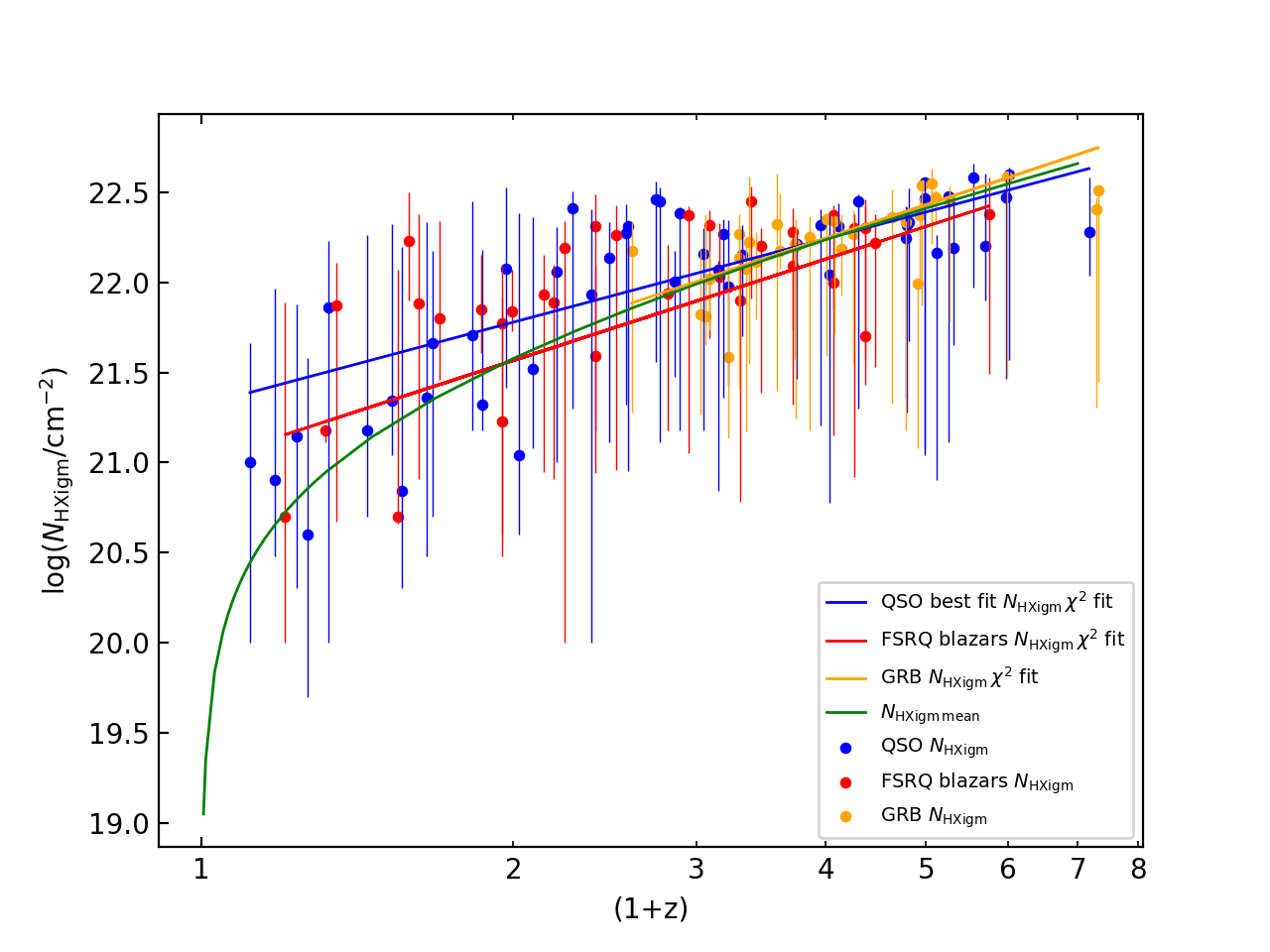} 
    \includegraphics[scale=0.55]{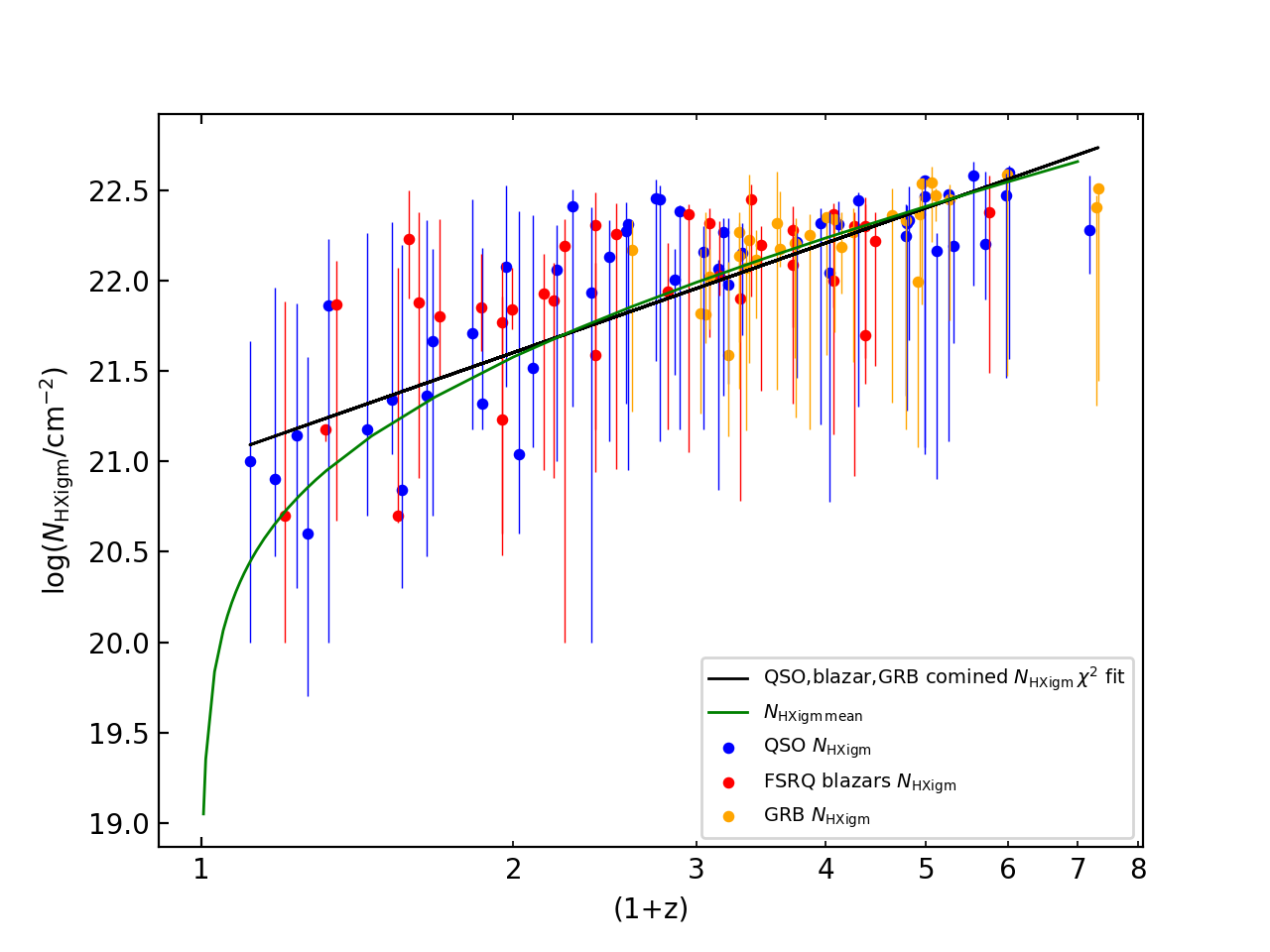}
    \end{tabular}
    \caption{$\mathit{N}\textsc{hxigm}$ versus redshift for the full QSO sample (blue) combined with the FSRQ blazars from D21b (red) and the GRBs (yellow) from D21a. In the left panel, each tracer group has its own $\chi^2$ line fit. The green line is the simple IGM model based on a mean IGM density of  n$_{0} = 1.7 \times 10^{-7}$ cm$^{-3}$ at $z = 0$ (see eq.\ref{eq:simpleIGM}). In the right panel,the $\chi^2$ line fit is for the entire tracer samples. }
        \label{fig:QSO_FSRQ_GRB_NHX_z}
\end{figure*}

\begin{table*}
    \renewcommand{\arraystretch}{1.3}
	\centering
	\caption{Summary results for the IGM parameters from the QSO, blazar and GRB samples from D21a and D21b, and this paper. The IGM parameters include the mean hydrogen density at $z = 0$ $n_0$ for the full redshift range and for $z > 1.6$, a  power  law  fit  to  the $\mathit{N}\textsc{hxigm}$ versus  redshift, mean temperature and metallicities, and the ranges.}
	    \label{tab:Alltracers}
	\begin{tabular}{cc@{\hspace*{0.7cm}} c@{\hspace*{0.5
	cm}}c@{\hspace*{0.5
	cm}}c@{\hspace*{0.5cm}}c@{\hspace*{0.5cm}}c@{\hspace*{0.5
	cm}}}
		\hline

		Tracer & QSO & Blazar & GRB & All \\
		\hline

        & & mean hydrogen density at $z = 0$ ($\times 10^{-7}$ cm$^{-3}$) & &   \\
        \hline
       Full redshift range & $2.8\pm{0.3}$ & $3.2\pm{0.5}$ & $1.8\pm{0.2}$ & \\
       $z > 1.6$ & $2.1\pm{0.3}$ & $2.1\pm{0.2}$ & $1.8\pm{0.2}$ & $2.0\pm{0.4}$\\
       \hline
      & &power  law  fit  to  the $\mathit{N}\textsc{hxigm}$ versus  redshift  & & \\ 
      \hline
      Slope index & $1.5\pm{0.2}$ & $1.8\pm{0.2}$ & $1.9\pm{0.2}$ & $2.0\pm{0.1}$\\
      \hline
       & &Temperature (log$(T/K))$ & & \\
       \hline
      Mean &$6.5\pm{0.1}$ & $6.1\pm{0.1}$ & $6.3\pm{0.2}$ & $6.3\pm{0.3}$ \\
      Range & $4.9 - 8.0$ & $5.0 - 8.0$ & $5.0 - 7.1$ & \\
      \hline
       & &Metallicity $[X/$H$]$ & & \\
       \hline
      Mean & $-1.3\pm{0.1}$ & $-1.6\pm{0.0}$ & $-1.8\pm{0.1}$ & $-1.5\pm{0.1}$\\
      Range & $-2.85$ to $-0.8$ & $-3.0$ to $-0.08$ & $-1.75$ to $-1.0$ & \\

      \hline
	\end{tabular}
\end{table*}

Intervening lensing galaxies on the QSO LOS have been observed over the years. We reviewed literature and identified 2 of our sample QSOs with confirmed intervening lenses, 042214.8-384453.0 and 100434.91+411242.8. The neutral column through these lensing galaxies was estimated as log($\mathit{N}_{\mathrm{H}\/\ \textsc{i}}$/cm$^{-2}) < 20$, two orders of magnitude lower than our measured $\mathit{N}\textsc{hxigm}$ \citep{Chen2012,carswell1996}.

In conclusion, our robustness tests have demonstrated that, with the possible exception of luminosity, we have ruled out alternative explanations for the observed $\mathit{N}\textsc{hxigm}$ redshift relation including reflection hump, soft excess, power law index, and spectral counts, intervening DLA and lensing galaxies. We note that the use of a log-parabolic power law may be showing an improved fit over a simple power law in all our QSO sample as either an intrinsic continuum feature, or or a slight signature of the reflection hump and soft excess. As for luminosity, in the next section we compare our QSO results with our previous GRB and blazar results to see if there are consistencies which would help rule out the luminosity degeneracy.

\section{Combined QSO, GRB and blazar sample analysis}\label{sec:combinedQSOGRBblazar}
In this section, we combine the results from our full series of papers on using different tracers to probe the IGM column density, temperature and metallicity. Given the differences in the tracer host environment, we adopted different approaches in estimating any host absorption. In D20 and D21a, we assumed that the GRB host intrinsic $\mathit{N}\textsc{hx}$ was  equal to the ionisation-corrected intrinsic neutral column measured in UV, using more realistic host galaxy metallicities, dust corrected where available. In D21b, using blazars, we assumed no host absorption, relying on the fact that blazars are thought to have negligible X-ray absorption on the LOS within the host galaxy due to the relativistic jet. Finally, in this paper using QSOs, as set out in Section \ref{sec:models}, we assume a CGM model absorption. Apart from these differences in modelling the host absorption, all other methods and models are consistently used for the three tracers.

In Fig. \ref{fig:QSO_FSRQ_GRB_NHX_z}, we plot the combined tracer samples for $\mathit{N}\textsc{hxigm}$ and redshift. In the left panel, the approximated linear $\chi^2$ fits are shown separately for each tracer. Though there are differences in the linear slopes, all three are reasonably close to the simple IGM curve. In the right panel, we show the $\chi^2$ linear fit for the combined samples. This fit is also close to the simple IGM curve. In Table \ref{tab:Alltracers}, we give the main IGM parameter results from each tracer and in combination including the mean hydrogen density at $z = 0$, $\mathit{N}\textsc{hxigm}$ versus  redshift power law fit, mean temperature and metallicities, and the ranges.

The first IGM parameter, the mean hydrogen density at $z = 0$ is given for both the full redshift range and also for our tracers with $z > 1.6$. Our GRB sample in D21a took $\mathit{N}_{\mathrm{H}\/\ \textsc{i}}$ data from \citet{Tanvir2019a} who had a cutoff at $z = 1.6$, as below this redshift, the observed Ly$\alpha$ transmission declines due to Earth's atmosphere. All of the values for $n_0$ are slightly higher than the simple IGM curve based on n$_{0}$ equal to $1.7 \times 10^{-7}$ cm$^{-3}$ (see Section \ref{sec:QSO results}). The overall mean across the three tracers for $z > 1.6$ is $2.0\pm{0.4}\times 10^{-7}$ cm$^{-3}$ which is consistent with the assumed density of the plotted IGM curve within the errors.  

The mean CIE IGM temperature across the tracers  is log($T/K) = 6.3\pm{0.3}$ with a full range from $4.9$ to $8.0$. The mean IGM metallicity across the tracers  is $[X/$H$] = -1.5\pm{0.1}$ with a full range from $-3.0$ to $-0.08$. These values are consistent with the CIE predictions for a warm/hot IGM. There is no apparent relation of temperature with redshift. 

We conclude that the consistent combined results of our samples demonstrate that the IGM is contributing to the absorption observed in the spectra, and that it is consistent with that predicted by the simple IGM model (eq.\ref{eq:simpleIGM}.) We caveat this conclusion noting that it is based on the assumption that the slab model, placed at half the tracer object redshift, is a reasonable representation of the LOS through the diffuse IGM.

\section{Discussion and comparison with other studies}\label{sec:Discuss}
Our work has found significant excess absorption (over our Galaxy and the QSO host) in QSO spectra. Excess X-ray absorption in QSOs has been reported in earlier studies,  predominantly in RLQ rather than RQQ \citep[e.g.][]{Elvis1994,Page2005,Reeves2000}. Initial possible explanations included the absorption being related to the jet, which was thought to be responsible for  the Doppler boosting  of the X-ray continuum \citep[e.g.][]{Reeves2000}. Most of these studies found the absorption tended to increase with QSO redshift. This would not support the jet absorption theory as the QSO jet luminosity was not found to increase with redshift \citep{Scott2011a}. \citet{Eitan2013} found that the optical depth increased with redshift for a sample of QSOs and GRBs, scaling as $(1 + z)^{2.2\pm0.6}$. This is very close to our combined tracer result for $\mathit{N}\textsc{hxigm}$ of $(1 + z)^{2.0\pm0.1}$. \citet{Eitan2013} postulated that their result could be explained by an ionized and clumpy IGM at $z < 2$, and a diffuse, cold IGM at higher redshift. This scenario was improved on by \citet{Starling2013}, who used a warm-hot absorber scenario for the IGM. \citet{Starling2013} concluded that their warm-hot IGM scenario could account for most of their estimated X-ray column density for GRB at $z > 3$ for IGM parameters log($T/K) = 5 - 6$ and $Z/Z\odot <0.2$. The main differences and caveats on the results of \citet{Eitan2013} and \citet{Starling2013}, are that they used the conventional assumption that all excess absorption is at the host redshift, despite dealing with IGM absorption on the LOS. Further, while \citet{Eitan2013} measured optical depth, \citet{Starling2013} used \textsc{absori} which was compared with \textsc{hotabs} for CIE modelling by D21a. \textsc{absori} is not self-consistent, and is limited to 10 metals fixed at solar metallicity except Fe \citep{Done1992}. D21a found \textsc{hotabs} to be superior for modelling a CIE IGM.

\citet{Campana2015} examined IGM absorption to GRBs and AGN using simulations. For GRBs, they reported log$(T/$K) $\sim  5 - 7$ and mean metallicity $Z = 0.03Z\sun$. In Section \ref{sec:combinedQSOGRBblazar}, we showed that our results across all our tracers, QSOs, blazars and GRBs are consistent for the IGM parameters. Our overall mean temperature and range for the IGM are log($T$/K) $= 6.5\pm{0.1}$, and $4.9 <$ log($T$/K) $< 8.0$. Our mean metallicity and range on solar units are 0.05Z\sun and 0.16Z\sun to 0.001Z\sun. These values are similar to \citet{Campana2015}.

\citet[hereafter A18]{Arcodia2018} used a blazar sample to investigate an IGM absorption scenario. Their IGM parameter results gave an average density ($z = 0$) of $n_0 = 1.0^{+0.53}_{-0.72} \times 10^{-7}$ cm$^{-3}$ and temperature log$(T/$K) $= 6.45^{+0.51}_{-2.12}$. The temperature is very close to our results. Our average density at $z = 0$ across all tracers, limiting the sample to $z > 1.6$ to accommodate the GRBs, is $n_0 = 2.0\pm{0.40} \times 10^{-7}$ cm$^{-3}$. A18's $n_0$ is less than the conventional simple IGM model of $n_0 = 1.7 \times 10^{-7}$ cm$^{-3}$ (see Section \ref{sec:QSO results}). However, we do note that in our Fig. \ref{fig:QSO_FSRQ_GRB_NHX_z}, some of the highest redshift tracers show $\mathit{N}\textsc{hxigm}$, with equivalent $n_0$ at $z = 0$ below the simple IGM curve assumption. A18's lower result could be explained by their use of conventional assumptions of neutral and solar absorption in the IGM and their use of an older \textsc{absori} based model. Alternatively, our combined results may indicate that a single uniform average density is over-simplistic across the full redshift range. The result is based on the homogeneity assumption and expansion of the Universe in the $\Lambda$CDM model. However, this does not factor in the structural changes and growth which are predicted to occur over redshift. For example, the fraction of matter in the IGM is expected to be much greater at higher redshift than lower redshift, as less matter had coalesced into galaxies and clusters \citep{McQuinn2016a}.

One of our assumptions is that of CIE. The relation between ionisation state and plasma temperature explicitly assumes that the gas is in an ionisation equilibrium \citep{Richter2008}. Opinions on the IGM equilibrium state have differed over the years \citep[e.g.][]{Branchini2009,Nicastro2018}. Plasma remains over-ionised at any temperature in non-equilibrium versus equilibrium conditions \citep{Gnat2007}. 
It is likely that a substantial part of the baryons in the Universe are located in low density regions where ionisation equilibrium conditions persist \citep{McQuinn2016a}. An underestimation of column density may arise due to assumed equilibrium conditions(D21a and references therein).

Generally, the fraction of RLQ to RQQ is $5 - 10\%$ and is possibly anti-correlated with redshift \citep[and references therein]{Rusinek-Abarca2021}. As RLQ tend to have far greater X-ray luminosity than RQQ, they are more frequently observed at higher redshift \citep{Worrall1987,Page2005}. In our study the fraction of RLQ is $\sim40\%$ reflecting the X-ray loudness bias due to our higher redshift range and choice of QSOs with high counts. However, if we look at the sample below $z < 3$ in Fig. \ref{fig:NHX_z} where the RQQ are dominant, and the RLQ fraction is $\sim26\%$, the redshift relation is still very clear indicating that the luminosity or high redshift RLQ bias is not driving the  $\mathit{N}\textsc{hxigm}$ redshift relation.

It is possible that some additional absorption occurs in the QSO host over and above our assumed CGM amount i.e. intrinsic dust or gas in the host galaxy inter-stellar medium. Alternatively, absorption could occur in the inter-cluster medium as many QSOs are located in galaxy clusters \citep{Elvis1994a}. However, higher absorption if related to neutral gas would result in higher dust measurements which are not observed \citep{Page2005}.

Significant curvature is present in the spectra of many low redshift QSOs, below $z < 1$. This fact, or that absorption features are not observed in such low redshift tracers has been used as an argument against IGM absorption \citep[and references therein]{Watson2012}. In our QSO sample, many of the lowest redshift QSOs closely follow the simple IGM curve. It is likely that spectral curvature is due to both intrinsic factors as well as IGM absorption, with the former dominant at low redshift, and the latter becoming dominant at higher redshift.

 Comparing with an alternative tracer type, Fast Radio Burst (FRB) dispersion measure (DM) is used to measure the total electron column density on the LOS to the FRB host.  The conventional approach with FRBs is to fix the host DM, scaled to reflect dispersion in the rest frame of the host \citep[e.g.][]{Shull2017,Macquart2020}. The assumption is then that all excess DM (over the host and our Galaxy) is due to the IGM, similar to our approach. This is contrary to the conventional approach with GRBs, blazars and QSOs where the assumption is all X-ray absorption in excess of our Galaxy is at the host redshift. Using FRBs, \citet{Macquart2020} derived a median baryon fraction of $\Omega_{b,H_0} = 0.056$ ($68\%$ confidence interval  $[0.046, 0.066]$). Based on this measurement, they conclude that their results are evidence of the missing baryons being present in the ionized intergalactic medium. Our median value for the baryon fraction for all our tracers with $z > 1.6$ is $\Omega_{b,H_0} = 0.048$  ($68\%$ confidence interval  $[0.039, 0.058]$  (derived from $n_0 = 2.0\pm{0.4} \times 10^{-7}$ cm$^{-3}$). For our full QSO sample only, the $\Omega_{b,H_0} = 0.068$, ($68\%$ confidence interval  $[0.061, 0.075]$) . These values are consistent with \citet{Macquart2020}.

\section{Conclusion}\label{sec:conclusion}

We used QSOs to probe the IGM column density, metallicity and temperature using a CIE model for the diffuse IGM. To isolate the IGM LOS contribution to the total absorption, we assumed that the QSO host absorption is based on a fixed model of CGM absorption.  We use the continuum total absorption as opposed to fitting individual absorption features as, currently, the required resolution is not available in X-ray.
  
 We tested our results for robustness covering: a relation between column density and spectral counts; spectral slope degeneracy with column density; reflection hump and soft excess impacts; luminosity column density relation, and any impact of large absorbers known of UV studies on the LOS.
  
 We aggregated our sample with the blazars from D21b and the GRB sample from D21a to present combined results for the IGM properties.
 
 Our main findings and conclusions are:
\begin{enumerate}
  \item The results for the IGM parameters are consistent across the GRBs, blazars and QSOs. The average results across the tracers for equivalent mean hydrogen density at $z = 0$ are $n_0 = 2.0\pm{0.4} \times 10^{-7}$ cm$^{-3}$ for $z > 1.6$. The combined results show similar values and correlation with redshift as the simple mean IGM density model, Fig. \ref{fig:QSO_FSRQ_GRB_NHX_z}, right panel. The $\mathit{N}\textsc{hxigm}$ versus redshift power law fit scales as $(1 + z)^{2.0\pm0.1}$.
  \item For our QSO sample in this paper, $n_0 = 2.8\pm{0.3} \times 10^{-7}$ cm$^{-3}$. The $\mathit{N}\textsc{hxigm}$ versus redshift power law fit scales as $(1 + z)^{1.5\pm0.2}$.
  \item The mean temperature across all the tracers for the CIE IGM is log($T/K) = 6.3\pm{0.3}$ with a full range from $4.9$ to $8.0$. The mean metallicity across the tracers for the CIE IGM is $[X/$H$] = -1.5\pm{0.1}$ with a full range from $-3.0$ to $-0.08$. These values are consistent with the CIE predictions for a warm/hot IGM. There is no evidence for evolution with redshift. 
  \item The mean temperature across our QSOs only sample from this paper for the CIE IGM is log($T/K) = 6.5\pm{0.1}$ with a full range from $4.9$ to $8.0$. The mean metallicity across the QSO sample for the CIE IGM is $[X/$H$] = -1.3\pm{0.1}$ with a full range from $-2.9$ to $-0.8$. These values are also consistent with the CIE predictions for a warm/hot IGM, and there is no evidence for evolution with redshift. 
  \item For the QSO sample, there is no obvious relation between $\mathit{N}\textsc{hxigm}$ and the robustness tested parameters for continuum power law index or spectral counts. Further, the possible effects of the reflection hump and soft excess were shown not to impact the results for $\mathit{N}\textsc{hxigm}$, and only improved the fit for the two lowest redshift QSOs. There was insufficient evidence for DLAs or intervening lens systems on the LOS to account for the measured $\mathit{N}\textsc{hxigm}$ for the QSOs. Finally, there is an apparent $\mathit{N}\textsc{hxigm}$ luminosity relation due luminosity bias in our sample, which is due to our sample selection for high counts and the dominance of RLQ at $z > 3$, which are more X-ray luminous than RQQ. Both $\mathit{N}\textsc{hxigm}$ and luminosity rise with redshift. However, the results for $\mathit{N}\textsc{hxigm}$ are consistent across all three tracers and this would support the argument that the QSO result for $\mathit{N}\textsc{hxigm}$ is not dominated or caused by luminosity.
  
 Overall in this series of papers D20, D21a and D21b, and this paper, we have demonstrated a consistent case for strong absorption in the IGM on the LOS to three different tracer types, QSOs, blazars and GRBs. We have taken a careful approach to isolating absorption by our Galaxy and the tracer host, by examining the differing host environment conditions known to exist for the tracer types. We have also examined the possible contribution on the LOS due to large absorbers from UV QSO studies and have subjected our results to a series of robustness tests. 
 
 As we have demonstrated that there is substantial absorption in the IGM and the mean column density is related to redshift, using the conventional assumption that all excess absorption is in the tracer host, while investigating high redshift objects could lead to errors in deriving their properties including intrinsic absorption. Our results could also be used to test cosmological models through observations of IGM properties from these high redshift tracers. Our estimated IGM properties will be improved by instruments such as Athena, with higher energy resolution, lower energy threshold and larger effective areas in soft X-ray energies.

\end{enumerate}

\section*{Acknowledgements}
 We thank Ehud Behar for a detailed and insightful review of the paper. We would like to thank E. Lusso for making available the data from the SDSS DR14 and 4\textit{XMM-Newton}-DR9 cross-correlation dataset, and E. Nardini for advice on using \textsc{pexrav} and \textsc{zbbody} for QSOs. This work made use of data supplied by \textit{XMM–Newton}, an ESA science mission with instruments and contributions directly funded by ESA member states and NASA. S.L. Morris acknowledges support from STFC (ST/P000541/1).
This project has received funding from the European Research Council (ERC) under the European Union's Horizon 2020 research and innovation programme (grant agreement No 757535) and by Fondazione Cariplo (grant No 2018-2329).


\section*{Data Availability}
Data including spectra fit plots for all the QSO sample, and MCMC integrated probability plots are available on request - please contact Tony Dalton. The reduced  \textit{XMM-Newton} spectral files are available by contacting Efrain Gatuzz. \textit{XMM-Newton} EPIC spectra data is available at http://xmm-catalog.irap.omp.eu/sources. SDSS data is available at http://skyserver.sdss.org/dr17.



\bibliographystyle{mnras}
\bibliography{references.bib}




\bsp	
\label{lastpage}
\end{document}